\documentclass{emulateapj}

\shorttitle{Abundances and density structure of the inner circumstellar ring around SN 1987A}
\shortauthors{Mattila et al.}

\begin{document}



\title{Abundances and density structure of the inner circumstellar ring around SN 1987A}


\author{Seppo Mattila\altaffilmark{1,2}, Peter Lundqvist\altaffilmark{2},
Per Gr\"oningsson\altaffilmark{2}, Peter Meikle\altaffilmark{3},
Raylee Stathakis\altaffilmark{4}, Claes Fransson\altaffilmark{2,5},
Russell Cannon\altaffilmark{4}}
\altaffiltext{1}{Tuorla Observatory, Department of Physics \& Astronomy, University of Turku, V\"ais\"al\"antie 20, FI-21500 Piikki\"o, Finland.}
\altaffiltext{2}{Stockholm Observatory, Department of Astronomy, AlbaNova University Center, SE-106 91 Stockholm, Sweden.}
\altaffiltext{3}{Astrophysics Group, Blackett Laboratory, Imperial College London, Prince Consort Road, London.}
\altaffiltext{4}{Anglo-Australian Observatory, PO Box 296, Epping, NSW 1710, Australia.}
\altaffiltext{5}{Oskar Klein Centre, Department of Astronomy, AlbaNova University Center, SE-106 91 Stockholm, Sweden.}




\begin{abstract}
We present optical spectroscopic data of the inner circumstellar ring around SN 1987A from the 
Anglo-Australian Telescope (AAT) and the Very Large Telescope (VLT) between $\sim$1400 and 
$\sim$5000 days post-explosion. We also assembled the available optical and near-infrared line fluxes
from the literature between $\sim$300 and $\sim$2000 days. These line light curves were fitted with a photoionization model
to determine the density structure and the elemental abundances for the inner ring. We found densities
ranging from 1$\times$10$^{3}$ to 3$\times$10$^{4}$ atoms cm$^{-3}$ and a total mass of the ionized gas
of $\sim$5.8 $\times$ 10$^{-2}$ M$_{\odot}$ within the inner ring. Abundances inferred from the optical and near-infrared data were 
also complemented with estimates of Lundqvist \& Fransson (1996) based on ultraviolet lines.
This way we found an He/H-ratio (by number of atoms) of 0.17 $\pm$ 0.06 which is roughly 30\% lower than previously estimated
and twice the solar and the Large Magellanic Cloud (LMC) value. We found an N/O-ratio of 1.5 $\pm$ 0.7, and the total (C+N+O)/(H+He)
abundance about 1.6 times its LMC value or roughly 0.6 times the most recent solar value. An iron abundance of 0.20 $\pm$ 0.11 times solar 
was found which is within the range 
of the estimates for the LMC. We also present late time ($\sim$5000 - 7500 days) line light curves of [O~III], [Ne~III], [Ne~IV], 
[Ar~III], [Ar~IV], and [Fe~VII] from observations with the VLT. We compared these with model fluxes and found that an additional $10^{2}$ atoms cm$^{-3}$
component was required to explain the data of the highest ionization lines. Such low density gas is expected in 
the H II-region interior to the inner ring which likely extends also to larger radii at higher latitudes (out of the ring plane). At epochs later than $\sim$5000 
days our models underproduce the emission of most of these lines as expected due to the contribution from the interaction of the supernova
ejecta with the ring.
\end{abstract}





\section{Introduction}

Supernova (SN) 1987A, with its surrounding triple ring structure is one of the best known 
astronomical images from the 1990's. However, the origin of these spectacular rings is still 
somewhat uncertain. It is believed that the material in these rings was ejected by the 
SN progenitor star $\sim$20,000 years ago. Also, the elemental abundances derived for the inner ring 
({\it e.g.}, Lundqvist \& Fransson 1996 hereafter LF96) strongly support a stellar origin for the ring 
material. Therefore, studies of the ring structure can also provide additional information on the 
progenitor star of SN 1987A which has been shown by detailed pre-explosion observations to have been
initially a 14-20 M$_{\odot}$ star and a blue supergiant of B3I spectral type at the time of the explosion (for review and
references see Smartt et al. 2009). Recently, binary merger models ({\it e.g.}, Morris \& Podsiadlowski 2007) 
have proposed to explain the blue supergiant nature of the SN progenitor as well as the formation of
the triple ring system. A rapidly rotating single star progenitor has been suggested as an alternative 
origin for the ring structure ({\it e.g.}, Chita et al. 2008).

The gas in the ring structure was photoionized by the extreme UV and soft X-ray SN flash with the first narrow optical and 
UV emission lines detected a few months after the explosion (Wampler et al. 1988; Fransson et al. 1989). The gas then cooled
and recombined until the start of the collision between the expanding SN ejecta and the material of the inner ring.
The first evidence that the ejecta/ring impact had started came from radio (Staveley-Smith et al. 1992)
and X-ray emission (Gorenstein, Hughes \& Tucker 1994) in mid-1990. The first signs at optical wavelengths were detected in 
Hubble Space Telescope (HST) images from 1995 (Lawrence et al. 2000). Recent high resolution X-ray images (Ng et al. 2009) from 2008 
indicate that the SN blast wave has already overtaken the inner ring. The UV-optical emission line fluxes from the inner ring have been modeled by LF96 up to $\sim$2000 days from the explosion. They found gas densities ranging from 6 $\times$ 10$^{3}$ cm$^{-3}$ 
to 3.3 $\times$ 10$^{4}$ cm$^{-3}$ and derived the following relative (ratio by number of atoms) abundances: 
He/H = 0.25 $\pm$ 0.05, N/C = 5.0 $\pm$ 2.0 and N/O = 1.1 $\pm$ 0.4 with an overall metal ({\it i.e.} C, N, and O) abundance 
of 0.30 $\pm$ 0.05 times solar. HST data up to $\sim 3500$ days were analyzed by Lundqvist \& Sonneborn (1997) and showed evidence 
of densities in the ring down to $\sim$2 $\times$ 10$^{3}$ cm$^{-3}$. Abundance estimates for metals in the SN 1987A CSM have recently also been made from X-ray observations (Zhekov et al. 2006, Dewey et al. 2008, Heng et al. 2008, Zhekov et al. 2009, 2010) of 
emission originating from the interaction between the expanding SN ejecta and the material in the inner ring. These observations 
yielded a similar N/O abundance as found by LF96, although the absolute C+N+O abundance is lower, and they also add 
abundance estimates of heavier metals.

The main aim of the present study is to investigate the density structure and the elemental abundances
of the inner ring using optical emission line fluxes from the entire era before the collision between the SN ejecta and the ring
material started to dominate these fluxes. To do this, we obtained optical spectra of the SN 1987A CSM 
with the Anglo-Australian Telescope (AAT) between $\sim$1400 and $\sim$4300 days and with the Very Large Telescope (VLT) 
between $\sim$5000 and $\sim$7500 days from the explosion. We also collected all the available optical/near-IR absolute line fluxes of the circumstellar 
ring from the literature for epochs between $\sim$300 and $\sim$2000 days. This data-set covers epochs between 
300 days and $\sim$7500 days from the SN explosion {\it i.e.} the entire era from the time when the emission from the inner ring 
first became visible at optical and near-IR wavelengths until and beyond the first signs of the collision between the SN ejecta and the inner
ring were detected at these wavelengths. Preliminary studies of the line light curves between $\sim$300 and $\sim$4300 have already been 
presented in Mattila (2002) and Mattila et al. (2003). The late time VLT observations have also been reported in Gr\"oningsson et al. (2008a,b).
In Section 2 we describe the AAT and VLT observations, data reductions and the available line fluxes from the literature. In Section 3,
the line light curves are modeled with a photoionization model to estimate the
density structure and elemental abundance of the inner ring. Finally, in Section 4
we discuss the results and in Section 5 present conclusions from this study.

\section{Observations}

\subsection{AAT Spectroscopy}
Optical spectroscopic observations of SN 1987A were carried out using the 
Royal Greenwich Observatory (RGO) Spectrograph on the AAT between 1991 and 
1998 (for full details of the AAT observations see Table 1). Data were obtained
at four different epochs: 1416, 1680, 1991, and 4309
days post-explosion. At each epoch, the observations comprised a brief, wide
slit integration, and a longer duration, narrow slit integration (see col. 6 in Table 1)
with spectral resolutions of FWHM $\sim$300-600 km~s$^{-1}$ and $\sim$150 km~s$^{-1}$, respectively.
The SN observations spanned air masses $\sim$1.5 to $\sim$3.0, making accurate flux calibration
quite challenging. We therefore carried out the observations with 
the slit position angle (PA, col 5 in Table 1) set to be roughly the same as the line joining 
Stars 2 and 3 (see Fig. 1). This meant that Star 2 lay well within the wide slit and so 
could be used to correct for the variable transmission of the atmosphere.  
However, an additional problem was that, since the slit PA was generally not 
at the parallactic angle, when the narrow slit was used, atmospheric refraction 
could introduce wavelength-dependent vignetting ({\it e.g.}, Filippenko 1982)
of the ring spectra. Moreover, the magnitude of this effect differed from that experienced by Star 2 since 
the latter lay nearer the edge of the slit. Consequently, the fluxing 
uncertainty introduced could be as large as $\pm$40$\%$. 

\begin{figure}
\centering
\vspace{+0.5cm}
\includegraphics[height=8cm]{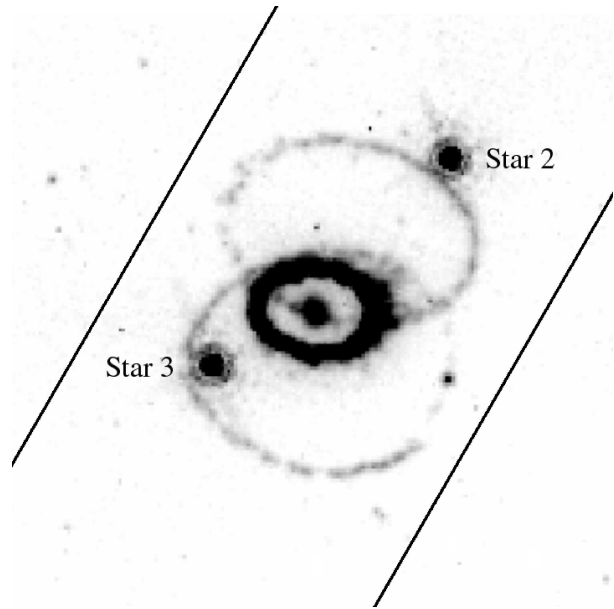}
\vspace{+0cm}
\caption{HST WFPC2 F656N image of SN 1987A ring structure from 10th July 1997
(3790 days). The position angle (-60$^{\circ}$) and width (5.3'' corresponding
to 1.3 pc at the distance of the LMC) of the wide slit used for the AAT observations 
on Dec. 1994 are indicated. North is up and East is to the left.}
\label{fig:1}
\vspace{+0.5cm}       
\end{figure}

Data reduction was carried out using standard {\tt IRAF} routines and for the 2D
frames included the overscan correction, bias subtraction, flat fielding (to correct for fringing)
and removal of cosmic rays. The background subtraction was performed using the {\tt IRAF BACKGROUND} 
task in two steps, first to remove the emission of atmospheric lines and then the more
complex Large Magellanic Cloud (LMC) background. The wavelength calibration was performed for the background
subtracted 2D frames using CuAr arc lamp spectra observed together with the SN.
An unweighted extraction was performed for the ring spectra because under the AAT 
seeing conditions the spectra of the two nearby stars were always blended with the 
spectrum of the ring making the use of any weighting algorithms unreliable.
For the wide-slit observations, extraction apertures were large enough to include the 
flux from both the two nearby stars and the inner ring (see Fig. 1) {\it i.e.} all the ring flux was included. For the
narrow slit observations smaller extraction apertures were used in order to maximise the S/N of the 
ring spectra and to avoid contamination by the strong stellar absorption lines in Stars 2 and 3
prominent at wavelengths shorter than $\sim$5000~{\AA}. The telluric correction was performed using 
spectra of the low metallicity F5 star HD 26169 located close to SN 1987A on the sky. This star has only a very 
few stellar features apparent in its spectrum which were removed by interpolation.
One of the spectrophotometric standards EG 21, LTT 1788, LTT 2415 or LTT 3218 (Hamuy et al. 1994)
was observed with each of the SN observations and was used for the relative flux calibration of the SN spectra.
Examples of the reduced narrow-slit spectra are shown in Fig. 2.

The SN observations spanned airmass values between $\sim$1.5 and $\sim$3.0. At such high airmasses
some of the observations are likely to be affected by clouds if the night is not completely clear.
Therefore, each wide slit spectrum was checked for the effects of attenuation by clouds, and a
correction applied to the affected spectra. This was carried out by making use of
the nearby Star 2 (for full details see Mattila 2002). The spectrum of Star 2 was extracted from each of the wide slit observations and
its continuum was fitted with a low order polynomial. The continuum with the highest counts (which also
had the expected shape for a stellar continuum) was selected to represent the unattenuated Star 2 continuum.
To compensate also for possible uncertainties in the relative fluxing a correction curve was created for each wide slit 
spectrum by dividing the 'real' Star 2 continuum by
the observed one. The extracted wide slit SN spectra were then multiplied by the corresponding correction curves.
In the AAT spectra the narrow lines arising from the inner ring were interactively fitted with a Gaussian profile on top
of the broad SN ejecta lines. As these lines have very different widths ({\it e.g.}, see Fig. 2) the SN ejecta lines could be treated as a 
local background for the narrow CSM lines and fitted with a low order polynomial on the two sides of each narrow 
line. We estimate that the uncertainties in the absolute fluxes measured from the wide slit spectra are
$\sim$$\pm$15\%. For the narrow slit spectra there is also another source of uncertainty in the relative fluxes
due to the differential refraction of light in the Earth's atmosphere. The flux calibration of the
narrow slit spectra and estimation of the associated uncertainties are described in Sect. 2.2.

\setlength{\tabcolsep}{0.03in}
\begin{table}
\vspace{0cm}
\hspace{-1cm}
\caption{Log of the AAT spectroscopy}
\label{1b}
\rotatebox{0}{
  \begin{tabular}{@{}lcccccccccccc@{}} \hline\hline
Date	   & Epoch & Wavelength & Exp.    & PA & Slit width          & Seeing  \\ 
           & [days] & [{\AA}]   & [sec] &[$\degr$]& [$\arcsec$]& [$\arcsec$] \\ \hline
1991       & 1416	& 5030-6300        & 2000	 & -65         & 6.7            & 2.0 \\
Jan 9-10   &            & 6175-7450        & 1000        & -65         & 6.7            & 1.8 \\
           &            & 7130-9900        & 1000       & -65         & 6.7            & 2.0 \\
           &            & 5030-6300        & 5400       & -65         & 2.0            & 2.0 \\
           &            & 6175-7450        & 6400       & -65         & 2.0            & 2.0 \\
           &            & 7130-9900        & 6900       & -65         & 2.0            & 2.0 \\ \hline 
1991       &  1680      & 6509-7790        & 300        & -42         & 6.7            & 2.7 \\
Oct 1      &            & 7600-8900        & 300        & -42         & 6.7            & 2.7 \\
           &            & 5400-6680        & 300        & -42         & 6.7            & 2.6 \\
           &            & 4314-5576        & 300        & -42         & 6.7            & 2.6 \\
           &            & 3220-4470        & 300        & -42         & 6.7            & 2.4 \\ 
           &            & 3220-4470        & 3600       & -42         & 2.0            & 2.4 \\
           &            & 4314-5576        & 3600       & -42         & 2.0            & 2.6 \\
           &            & 5400-6680        & 3600       & -42         & 2.0            & $\sim$2 \\
           &            & 6509-7790        & 3600       & -42         & 2.0            & 2.7 \\
           &            & 7600-8900        & 3600       & -42         & 2.0            & $\sim$2 \\ \hline
1994       & 2864    & 3200-4780           & 1000       & -60         & 5.3            & 1.1 \\
Dec 27-28  &            & 5980-7600        & 800, 100   & -60         & 5.3            & 1.1 \\
           &            & 4600-6200        & 1000       & -60         & 5.3            & 1.1 \\
           &            & 7380-9000        & 600        & -60         & 5.3            & 1.4 \\
           &            & 3200-4780        & 10800      & -60         & 1.5            & 1.1 \\
           &            & 4600-6200        & 10800      & -60         & 1.5            & 1.1 \\
           &            & 5980-7600        & 9000       & -60         & 1.5            & 1.1 \\ 
           &            & 7380-9000        & 7600       & -60         & 1.5/1.7        & 1.4 \\ \hline
1998       & 4309    & 3250-6370           & 600        & -60         & 10.0           & 1.1 \\
Dec 10-11  &            & 5180-8320        & 600        & -60         & 10.0           & 1.1 \\
           &            & 7850-11040       & 4400       & -60         & 10.0           & $\sim$3 \\ 
           &            & 3250-6370        & 7200       & -60         & 2.0            & 1.9 \\
           &            & 5180-8320        & 1200       & -60         & 2.0            & 1.1 \\
           &            & 6000-7600        & 8000       & -60         & 2.0            & 2.2 \\ \hline
\label{AAT2}
\end{tabular}}
\end{table}

\subsection{Absolute Flux Calibration with Archival HST Data}
The AAT observations were obtained with the seeing ranging between $\sim$1" and $\sim$3" and so 
the inner ring spectra were always significantly contaminated by light from 
Stars 2 and 3 located 2.9'' and 1.6'', respectively, from the SN (see Fig.~1).
In general, the CSM emission lines could be easily distinguished from the continuum originating from the SN and the two 
stars.  However, Star 3 is a Be star showing strong H$\alpha$ and H$\beta$ 
emission together with optical variability of $\sim$0.5 magnitudes (Wang et al. 1992).  
Consequently, it could affect significantly the observed line fluxes from the 
inner ring, especially at the later epochs when the ring was fainter.  To 
assess and correct for possible Star 3 contamination, we searched the HST 
archive for suitable optical spectroscopic (STIS) and photometric (WFPC2) 
observations (see Table 2) covering the epochs of our AAT observations.

\begin{table}
  \caption{HST observations of Star 3 and the inner ring}
  \begin{tabular}{@{}lccccccc@{}} \hline\hline
Date       & Epoch~ & Instr.~~     & \multicolumn{3}{c}{Star 3} & \multicolumn{2}{c}{Inner ring} \\
           & (days)&            & H${\alpha}$~    & H${\beta}$~    & m(B)~~  & F656N~    & F502N~ \\ \hline
1994 Feb 3 & 2537  & WFPC2      & 39             & ...           & 15.97 & 226 (244)& 57 (70)\\ 
1994 Sep 24& 2770  & WFPC2      & ...            & ...           & 15.96 & ...      & 48 (50)\\
1996 Feb 6 & 3270  & WFPC2      & ...            & ...           & 15.81 & ...      & 38 (41)\\
1997 Apr 26& 3715  & STIS       & 41             & 12            & ...   & ...      & ...\\
1997 Jul 10& 3790  & WFPC2      & 45             & ...           & 16.03 & 209 (167)& 33 (39)\\
1997 Dec 9 & 3942  & STIS       & 46             & 13            & ...   & ...      & ...\\       
1998 Feb 5 & 4000  & WFPC2      & 46             & ...           & ...   & 175 (157)& ... \\
1999 Apr 21& 4431  & WFPC2      & 52             & ...           & ...   & 166      & ...\\
2000 May 1 & 4816  & STIS       & 36             & ...           & ...   & ...      & ... \\ \hline\\
\end{tabular}
\label{STIS2}
\noindent
The H$\alpha$ and H$\beta$ fluxes and $B$-band magnitudes of Star 3 and F656N and F502N fluxes
of the inner ring as obtained from the HST observations. The line fluxes for H$\alpha$
(after correcting for Star 3) + [N II] $\lambda$6548 and [O III] $\lambda$5007 as obtained for 
the inner ring from the AAT observations and interpolated to the epochs of the HST observations 
are given for comparison in brackets after the F656N and F502N fluxes, respectively. All the fluxes are in the 
units of 10$^{-15}$ ergs$^{-1}$cm$^{-2}$.
\end{table}

\begin{figure*}
\centering
\vspace{+0.5cm}
\includegraphics[height=15cm,angle=-90]{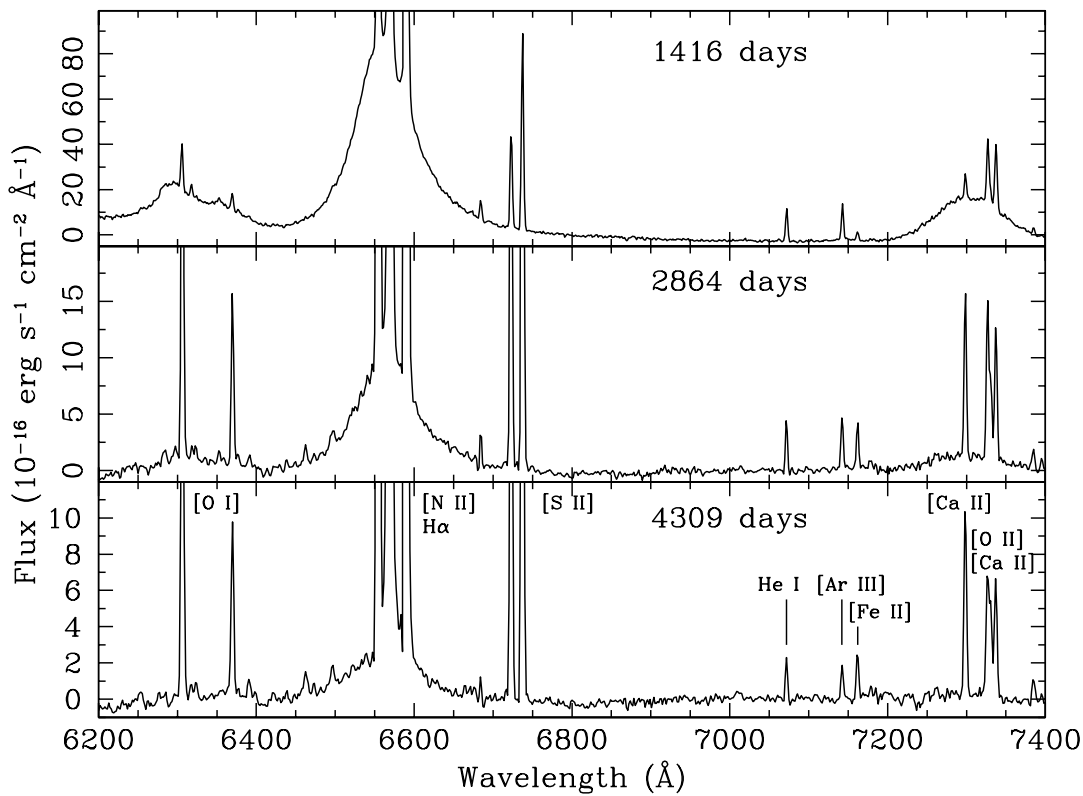}
\vspace{-0cm}
\caption{Continuum subtracted narrow-slit AAT spectra of SN~1987A and its ring structure showing the spectral region
between 6200~\AA~and 7400~\AA~on three epochs, Jan. 1991 (1416 days), Dec. 1994 (2864 days), and Dec. 1998 (4309 days).
The scale of the plots has been selected to better show the faint lines measured from these narrow slit spectra
whereas the brightest lines (measured from the wide slit spectra) extend outside the plotting range. The spectral
lines included in this study have been identified.}
\label{fig:1}
\vspace{+0.5cm}       
\end{figure*}

STIS spectra with a large enough slit aperture (52"x2") and suitable centering 
and orientation to include Star 3 were selected from the HST archive
at three epochs of observation.
Pipeline reduced STIS data were acquired from the HST archive and tasks from the {\tt IRAF} 
{\tt STSDAS} package were used to combine the pipeline-reduced single exposures, rejecting for 
cosmic rays and hot pixels. The emission lines from the sky were subtracted using the {\tt IRAF 
BACKGROUND} task with a 1st or 2nd order polynomial fit to the sky lines. However, the
subtraction of the background emission originating from spatially extended structures such as 
the outer rings was not possible and therefore, some faint nebular background emission is 
visible in the extracted Star 3 spectra. The Star 3 spectra were extracted using the {\tt STSDAS 
X1D} task with the recommended aperture size of 5 pixels (0.25$\arcsec$) selected to match the size 
of the stellar PSF. This task traces the center of the spectrum by cross correlation and automatically 
performs a 2-D wavelength calibration and flux calibration for the data. 
The only strong lines visible in the reduced Star 3 spectra are H$\alpha$ and H$\beta$ confirming the findings 
of Wang et al. (1992) that there are indeed no forbidden lines present in the Star 3 spectrum.

\begin{table*}
\begin{center}
\vspace{-0.4cm}
\hspace{0cm}
\caption{Line fluxes of the inner ring from the literature (307-2122 days).}
\label{1a}
\rotatebox{0}{
  \begin{tabular}{@{}lcccccccccccccccccccc@{}} \hline\hline
line	&  	   	 ~307 & 511  & 552   & 574   & 586   & 668   & 678   & 695   &734/735& 840   & 958 & 1050 & 1114 & 1280 & 1344 & 1348 & 1469 & 1734 & 1822 & 2122 \\ \hline
H$\alpha$ &	   	 ...  & ...  & ...   & ...   & 283   & ...   & 269   & ...   & 324   & ...   & ... & 251  & ...  & 143  &  140 & ...  & ...  & ...  & ...  & ...   \\ 
H$\beta$	&		  ~17& 46    & 64    & ...   & 72    & ...   & 74    & ...   & 100   & ... & ...  & 69   & ...  & 42   &  39  & ...  & ...  & ...  & ...  & ...\\ 
H$\gamma$ &		  ~10 & ...  & 27    & ...   & 35    & ...   &  24   & ...   &  21   & ...   &...  & 29	& ...  & 14   &  16  & ...  & ...  & ...  & ...  & ...    \\ 
H$\delta$ &		  ~25 & ...  & ...   & ...   & 26    & ...   &  8    & ...   &  14   & ...   & ... & ...	& ...  & 8.8  &  9.1 &  ...  & ...  & ...  & ...  & ...\\ 
H(Pa$\gamma$) &           ... & ...  & ...   & ...   & ...   & ...   & ...   & ...   & ...   & ...   & 35  &...	& 50   & ...  & ...  &   28 & 10 & 12.4 & 9.5 & 6.0 \\  
H(Pa$\beta$) &	          ... & ...  & ...   & ...   & ...   & ...   & ...   & ...   & 50    &  40   & 90  &...	& 70   & ...  & ...  &   34 & 10 & 9.0  & 7.0 & 6.5 \\ 
H(Br$\gamma$) &           ... & ...  & ...   & ...   & ...   & ...   & ...   & ...   & ...   & ...   & ... &...	&...   & ...  & ...  & ...  & 5.5& 1.5  & 2.0 & 1.5 \\  \hline
He II $\lambda$4686 &		  ... & ...  & 24    & ...   & 19    & ...   &  14   & ...   &  24   & ...   & ... & 14	&...   & 5.0  &  6.8 & ...  & ...  & ...  & ...  & ...  \\ \hline      
He I $\lambda$5876 &		  ... & ...  & ...   & ...   & 17    & ...   &  19   & ...   &  15   & ...   & ... & 16	&...   & 7.1  &  7.0 & ...  & ...  & ...  & ...  & ... \\  
He I $\lambda$7065 &		  ... & ...  & ...   & ...   & ...   & ...   & ...   & ...   & ...   & ...   & ... &...	&...   & 1.8  & ...  & ...  & ...  & ...  & ...  & ...  \\ 
He I $\lambda$10830 &		  ... & ...  & ...   & 1400  & ...   & 2400  & ...   & 2200  & 1500  & 1000 & 1400&...	&   650& ...  & ...  &   390 & 360 & 120 & 90 & 53\\ 
He I $\lambda$20581 & 		  ... & ...  & ...   & ...   & ...   & ...   & ...   & ...   & ...   & ...   & 5   &...	&   5  & ...  & ...  & ...  & 7.0 & 3.5 & 3.0 & 2.0 \\  \hline      
$[$N II$]$ $\lambda\lambda$6548,6583 &  ... & ...  & ...   & ...   & 792   & ...   &  796  & ...   &   938 & ...   & ... & 842	&...   & 286  &   349 & ...  & ...  & ...  & ...  & ...  \\
$[$N II$]$ $\lambda$5755 &	  ... & ...  & ...   & ...   & 20    & ...   &  24   & ...   &   23  & ...   & ... & 20	&...   & 9.5  &   8.0 & ...  & ...  & ...  & ...  & ...\\ \hline
$[$O I$]$ $\lambda\lambda$6300,6364 &	  ... & ...  & ...   & ...   & ...   & ...   & ...   & ...   &   11  & ...   & ... &...	&...   & 3.6  & ...   &  ...  & ...  & ...  & ...  & ...\\        
$[$O II$]$ $\lambda\lambda$3726,3729 &  ... & ...  & ...   & ...   & ...   & ...   & ...   & ...   &   77  & ...   & ... & 75	&...   & 71   &   49  & ...  & ...  & ...  & ...  & ...\\ 
$[$O II$]$ $\lambda\lambda$7319,7331 &  ... & ...  & ...   & ...   & ...   & ...   & ...   & ...   & ...   & ...   & ... &...	&...   & 12.9 & ...   & ...  & ...  & ...  & ...  & ... \\
$[$O III$]$$\lambda\lambda$4959,5007 &  ~433& ... & 973   & ...   & 857   & ...   &  668  & ...   &    825& ...   & ... & 393  &...   & 150  &  144  & ...  & ...  & ...  & ...  & ... 	\\ 
$[$O III$]$ $\lambda$4363 &	   ~27&  24  & 35    & ...   & 42    & ...   &  24   & ...   & ...   & ...   & ... & 18	&...   & 5.5  &    5.9& ...  & ...  & ...  & ...  & ...  \\ \hline
$[$Ne III$]$ $\lambda$3869      & ... & ... & ...   & ...   & ...   & ...   & ...   & ...   &    69 & ...   & ... & 60	&...   & 18   &   19  & ...  & ...  & ...  & ...  & ...  \\ \hline
$[$S II$]$ $\lambda\lambda$6716,6731 &   ... & ... & ...   & ...   & ...   & ...   &  36   & ...   &    41 & ...   & ... & 38	&...   & 17   &    20 & ...  & ...  & ...  & ...  & ... \\ 
$[$S II$]$ $\lambda\lambda$4069,4076 &   ... & ... & ...   & ...   & 24    & ...   &  15   & ...   &    15 & ...   & ... &...	&...   & 4.5  &  4.1  & ...  & ...  & ...  & ...  & ... \\ 
$[$S III$]$ $\lambda\lambda$9069,9532 &   ...& ... & ...   & ...   & ...   & ...   & ...   & ...   & ...   & ...   & ... &...	&...   & ...  & ...   & ...  & 61   & 40   & ...  & ... \\ \hline
$[$Fe II$]$ $\lambda$5159     &   ... & ... & ...   & ...   & ...   & ...   & ...   & ...   & ...   & ...   & ... &...	&...   & 0.8  & ...   & ...  & ...  & ...  & ...  & ... \\  
$[$Fe II$]$$\lambda$7155     &   ... & ... & ...   & ...   & ...   & ...   & ...   & ...   & ...   & ...   & ... &...	&...   & 0.8  & ...   & ...  & ...  & ...  & ...  & ... \\ 
$[$Fe II$]$$\lambda$12567   &   ...  & ... & ...   & ...   & ...   & ...   & ...   & ...   & ...   & ...   & ... &...	&...   & ...  & ...   & ... & 6.0   & 4.5  & 3.0  & 2.0 \\ 
$[$Fe II$]$ $\lambda$16435    &   ... & ... & ...   & ...   & ...   & ...   & ...   & ...   & ...   & ...   & ... &...	&...   & ...  & ...   & ... & 9.0   & 4.0  & 3.0  & 1.5 \\ 
$[$Fe III$]$$\lambda$4658    &   ... & ... & ...   & ...   & ...   & ...   & ...   & ...   & ...   & ...   & ... &...	&...   & 1.9  & ...   & ...  & ...  & ...  & ...  & ... \\ \hline
\label{STIS}
\end{tabular}}
\end{center}
\vspace{-0.5cm}
\noindent \\
The fluxes are in units of 10$^{-15}$ erg s$^{-1}$ cm$^{-2}$. 
The fluxes for 307 days were taken from Wampler $\&$ Richichi (1989), and were 
reddened back to observed fluxes using $E(B-V) = 0.05$ and 0.10 for the Galaxy and the LMC, respectively, 
and the extinction laws from Howarth (1983) assuming R$_{V}$ = 3.1. The fluxes for 511, 552, 586, 678, and 
735 days are from Wampler {\it et al.} (1989). The fluxes at 574, 668, 695, 734, 840, 
958, and 1114 days are from Meikle {\it et al.} (1993). The fluxes at 1050 days are from Menzies (1991), 
reddened back to the observed fluxes using $E(B-V) = 0.05$ and 0.19 for the
Galaxy and the LMC, respectively, and the extinction laws from Howarth (1983). 
The [O III] (4363, 4959, 5007) and [N II] (5755, 6548, 6583) fluxes given by Menzies (1991) are from day 1050 
whereas the other line fluxes are averages of the observed fluxes between $\sim$ 850 and 1200 
days. The fluxes at 1280 days are from Wang (1991). The fluxes at 1344 days are from Cumming (1994), 
reddened back to the observed fluxes using $E(B-V) = 0.05$ and 0.16 for Galaxy and the LMC, respectively 
and the extinction laws from Howarth (1983). The fluxes at 1348, 1469, 1734, 1822, and 2122 days 
are from Fassia {\it et al.} (2002).
\vspace{+0.5cm}
\end{table*}

WFPC2 images obtained using the F502N and F656N filters were selected from the HST archive 
at six epochs contemporary with our AAT spectra ({\it e.g.}, see Fig. 1). Pipeline reduced WFPC2 data were acquired 
from the HST archive and tasks from the {\tt IRAF} {\tt STSDAS} package were used to
combine the pipeline reduced repeat images rejecting for cosmic rays and hot pixels.
Aperture photometry was then performed for Star 3 using the recommended aperture diameter 
of 0.5$\arcsec$ and then corrected to an infinite aperture with the standard aperture correction.
As we are interested in the H$\alpha$ emission line flux on top of the stellar continuum, the narrow 
band fluxes of Star 3 needed to be corrected for the contribution from the continuum. The level of the continuum 
at the wavelength of H$\alpha$ was measured from the STIS spectra and was subtracted from the measured 
F656N flux densities. However, as Star 3 is known to vary with an amplitude of $\sim$0.5 magnitudes at 
optical wavelengths its continuum level can be expected to vary accordingly. Therefore, the continuum 
subtraction can be expected to introduce an error of about $\pm$20$\%$ to the H$\alpha$ emission line 
fluxes. The continuum subtracted H$\alpha$ fluxes are listed in Table 2.
As no forbidden emission lines have been seen in the Star 3 spectrum the F502N band
(centered on the [O III] line at 5007{\AA}) can be expected to contain pure continuum flux from the star.
Therefore, the narrow band fluxes observed with the F502N filter can be used to estimate the B-band 
magnitudes of Star 3. The magnitudes of Star 3 transformed to the Johnson B-band are listed in
Table 2. In the same way a magnitude of m$_{B}$ = 15.01 was obtained for Star 2 which is close to the 
earlier ground based measurement m$_{B}$ = 14.97 (Walborn {\it et al.} 1993).

Average H$\alpha$ and H$\beta$ fluxes of 4.4 and 
1.3 $\times$ 10$^{-14}$ erg s$^{-1}$ cm$^{-2}$, respectively, were found for Star 3, with a maximum 
variability of $\sim$20$\%$ over a 6 year time span. These fluxes were used to correct for 
Star 3 contamination of the hydrogen line fluxes in the wide slit AAT observations. 
The additional uncertainty in the H$\alpha$ and H$\beta$ fluxes due to the correction for the
Star 3 contamination was estimated to be $\pm$5\% for the 1416 and 1680 days epochs and
$\pm$15\% for the 2864 and 4309 days epochs.
The absolute flux calibration for the vignetted narrow slit spectra was obtained by scaling them
to match the unvignetted wide slit spectra using the brightest forbidden 
lines present in both the spectra. Deriving line fluxes for the whole inner ring 
this way assumes homogeneous ring geometry. Comparison to the line fluxes from the wide slit
spectra indicate an uncertainty of up to $\pm$40\% for the line fluxes measured from the
narrow slit spectra.

The accuracy of the absolute fluxing of our AAT observations were also checked using the 
HST inner ring data (see also Mattila 2002). The F502N filter is very useful for this as it contains the well isolated [O III] 
line at 5007~{\AA}. In addition, we used the F656N imaging data as this filter covers well the H$\alpha$ 
and the [N II] line at 6548~{\AA}, and is only slightly contaminated by the nearby [N II] line at 6583~{\AA}.
Therefore, the fluxes of the inner ring were measured from the F502N and F656N images using an aperture
with a 1.35$\arcsec$ radius, an ellipticity of 0.75, and a PA selected to match the orientation of
the ring in a given HST image. The flux from the SN in the middle of the ring was measured with
a circular aperture with a radius of 0.32$\arcsec$ selected to include most of the flux from the SN
avoiding the emission from the ring. The flux measured from the SN was typically $\sim$5-10$\%$ of 
the ring flux in the F656N images. The fluxes of the inner ring, corrected for the SN emission, are listed
in Table 2 together with the corresponding inner ring fluxes as measured from the AAT spectra 
and interpolated (after the Star 3 correction) to the epochs of the HST observations. Comparison between
these fluxes shows the AAT observations to be well consistent with the HST measurements to within a maximum
deviation of 25\%.

\subsection{VLT Spectroscopy}
High resolution (FWHM $\sim$ 6 km~s$^{-1}$) spectroscopic observations were obtained with
the Ultraviolet and Visual Echelle Spectrograph (UVES) at the ESO 
VLT during 2000 December 9-14 ($\sim$5040 days post-explosion),
2002 October 4-7 ($\sim$5703 days post-explosion), 2005 March 21 - 
April 12 ($\sim$6618 days post-explosion), 2005 October 20 - November
12 ($\sim$6826 days post-explosion), 2006 October 1 - November 15
($\sim$7183 days post-explosion), and 2007 October 23 - November 28
($\sim$7559 days post-explosion). The UVES data cover the spectral range 
$\sim$3100-10,000 \AA. The slit width was 0.8'' and oriented at a position angle of 30$^{\circ}$
across the SN. The airmass of these observations ranged between $\sim$1.4 and 1.8 and the seeing spanned
0.4-1.4''. The total exposure time was about 10,000 sec in each epoch. 
For full details of the observations see Table 1 in Gr\"oningsson et al. (2008b).
An atmospheric dispersion corrector was always used for these UVES observations
to minimize any wavelength dependent slit losses. The data reduction, flux 
calibration and the measurement procedure of the line fluxes are discussed in detail
in Gr\"oningsson et al. (2008a). With the relatively 
narrow slit only a fraction the ring flux was
covered by the observations. To estimate this fraction and to check the accuracy
of the flux calibration we compared the UVES fluxes to data taken with the
HST at a similar epoch. From this comparison
we estimated that the correction factor needed to compensate for the slit
losses was 2.4 (Gr\"oningsson et al. 2008b, Appendix A)
and the uncertainty of the absolute fluxes should be less
than $\sim$$\pm$20\% (see also Gr\"oningsson et al. 2008a).

\begin{table*}
\begin{center}
\vspace{0cm}
\hspace{-1cm}
\caption{Line fluxes of the inner ring from AAT and VLT (1416-7559 days)}
\label{1b}
\begin{tabular}{@{}lcccccccccc@{}} \hline\hline
line	&  	   	   ~1416     & 1680    & 2864     &4309     & 5040 & 5703    & 6618    & 6826    & 7183     & 7559 \\ \hline
H$\alpha$ &	   	    ~226$^{b}$& 186$^{b}$&86$^{b}$&76$^{b}$ & 55   & ...     & ...     & ...     & ...      & ...\\
H$\beta$	&          ...       & 47$^{b}$& 32$^{b}$&15$^{b}$  & 14   & ...     & ...     & ...     & ...      & ... \\
H$\gamma$ &		   ...       & 14$^{b}$&14       &8.4       & ...  & ...     & ...     & ...     & ...      & ...\\
H$\delta$ &		   ...       &  7.9    & ...     &3.3       & ...  & ...     & ...     & ...     & ...      & ...\\ \hline 
He II $\lambda$4686 &		   ...       &9.3     &5.7$^{b}$ &2.4       & ...  & ...     & ...     & ...     & ...      & ... \\        
He I $\lambda$5876 &		     ~12$^{b}$&15     &6.2$^{b}$ & 4.0      & 2.8  & ...     & ...     & ...     & ...      & ...\\
He I $\lambda$7065 &		    ~3.9$^{b}$&4.5    & 1.9$^{b}$&1.0$^{b}$ & 0.8  & ...     & ...     & ...     & ...      & ...\\  \hline
$[$N II$]$ $\lambda\lambda$6548,6583 &    ~633$^{b}$ & 580$^{b}$ & 515$^{b}$ & 340$^{b}$ & 226  & ...     & ...     & ...     & ...      & ...\\  
$[$N II$]$ $\lambda$5755 &	     ~15$^{b}$&15$^{b}$&10$^{b}$ &7.3$^{b}$ & 2.9  & ...     & ...     & ...     & ...      & ...\\ \hline
$[$O I$]$ $\lambda\lambda$6300,6364 &	    ~8.9$^{b}$&15.1    &18$^{b}$ &14$^{b}$  & 9.9  & ...     & ...     & ...     & ...      & ...\\       
$[$O II$]$ $\lambda\lambda$3726,3729 &    ...       &99      &70$^{b}$ &45        & 36   & ...     & ...     & ...     & ...      & ...  \\  
$[$O II$]$ $\lambda\lambda$7319,7331 &     ~22      &16      & 9.8     & 5.5      & 3.4  & ...     & ...     & ...     & ...      & ... \\  
$[$O III$]$ $\lambda\lambda$4959,5007&     ...      &189$^{b}$& 61$^{b}$&48       & 28   & 40      & 53      & 60      & 57       & 69 \\ 
$[$O III$]$ $\lambda$4363 &	    ...       &5.6$^{b}$&3.2     &2.1       & 1.4  & ...     & ...     & ...     & ...      & ... \\ \hline
$[$Ne III$]$ $\lambda$3869&        ...       &18      & 11      &12        & ...  & 5.2     & 8.5     & 8.5     & 7.8      & 6.9 \\ 
$[$Ne IV$]$ $\lambda\lambda$4714,4726     ...       & ...    & ...     & ...      & ...  & 0.11    & ...     &0.10     &0.16      &0.20\\ \hline
$[$S II$]$ $\lambda\lambda$6716,6731 &     ~40$^{b}$&36$^{b}$& 49$^{b}$&38$^{b}$  & 28   & ...     & ...     & ...     & ...      & ... \\ 
$[$S II$]$ $\lambda\lambda$4069,4076 &    ...       &5.6     & 9.2     &6.0       & 4.3  & ...     & ...     & ...     & ...      & ...\\
$[$S III$]$ $\lambda\lambda$9069,9532 &    ~51$^{b}$& ...    & ...     &14        & 5.0  & ...     & ...     & ...     & ...      & ...\\ \hline      
$[$Ar III$]$ $\lambda\lambda$7136,7751&   ~7.0      &7.7     & ...     & ...      & 1.2  & ...     & ...     & ...     & ...      & ...\\ 
$[$Ar IV$]$ $\lambda\lambda$4711,4740&   ...       & ...    & ...     & ...      & ...  & 1.3     &1.7      &1.9      &1.8       &1.6\\ \hline
$[$Ca II$]$ $\lambda\lambda$7291,7324 &    ~5.7     &6.8     &  8.0    &6.2       & 3.3  & ...     & ...     & ...     & ...      & ...\\ \hline
$[$Fe II$]$ $\lambda$5159 &	     ...      &0.9     & 1.0     &1.1       & 0.48 & ...     & ...     & ...     & ...      & ...  \\
$[$Fe II$]$ $\lambda$7155 &	    ~1.2      &1.5     & 1.9     &1.0       & 0.73 & ...     & ...     & ...     & ...      & ...\\ 
$[$Fe II$]$ $\lambda$7453 &	    ~0.6      &0.7     & 0.7     &0.2       & 0.24 & ...     & ...     & ...     & ...      & ...\\
$[$Fe III$]$ $\lambda$4658 &       ...       &2.2     &1.1      &1.2       & ...  & ...     & ...     & ...     & ...      & ...\\ 
$[$Fe VII$]$ $\lambda$5721 &       ...       & ...    & ...     & ...      & ...  &  0.14   &0.10     & 0.092   & 0.11     & 0.096 \\ \hline
\label{STIS}
\end{tabular}
\end{center}
\vspace{-0.5cm}
\noindent \\
The fluxes are in units of 10$^{-15}$ erg s$^{-1}$ cm$^{-2}$ and are all observed fluxes (not corrected for reddening)
from this study. The fluxes at 1416-4309 days are from the AAT and the later ones from the VLT.
The H$\alpha$ and H$\beta$ fluxes for these epochs have been corrected for the contamination by Star 3. The 
fluxes measured directly from the AAT wide slit spectra (marked with b) have a smaller uncertainty ($\pm$15 - 21\%) than the 
fluxes measured from the narrow slit spectra ($\pm$43\%). The fluxes at 5040-7559 days have uncertainties of $\pm$20\%
except [S II] $\lambda\lambda$9069,9532 for which the uncertainty is $\pm$50\%.
\end{table*}

\subsection{The Line Light Curves}
We assembled from the literature all the available optical/near-IR line fluxes of the inner ring between $\sim$300 and $\sim$2100 days.
Early time optical CSM line fluxes were collected from 
Wampler \& Richichi (1989), Wampler {\it et al.} (1989), Wang (1991), Menzies (1991), and Cumming (1994). Near-IR 
(NIR) fluxes covering both early and late time epochs were taken from Meikle {\it et al.} (1993), and  Fassia {\it et al.} 
(2002). The fluxes from Wampler \& Richichi (1989), Menzies (1991), and Cumming (1994) were reddened back 
to the observed fluxes to enable consistent dereddening for the data from all
the epochs. The fluxes given by Cumming (1994) had been dereddened using the extinction laws of
Howarth (1983) (R. Cumming, private communication). Without any additional information, the same extinction
laws were also used when reddening the fluxes given by Wampler \& Richichi (1989) and Menzies (1991).
In all cases it was assumed that the Galaxy contributed $E(B-V) = 0.05$ with the rest of the reddening (see the caption of Table 3) due to
the extinction within the LMC. Meikle {\it et al.} (1993) estimate errors of $\pm$15$\%$ for their He I $\lambda$10830 line fluxes. 
Also, Fassia {\it et al.} (2002) estimate errors of $\pm$35$\%$ for their IJ-band line fluxes. However, the other authors do not 
give any error estimates for their measurements. We therefore adopted errors of $\pm$30$\%$ for these early time optical line fluxes.
The fluxes from the literature are listed in Table 3.

\begin{center}
\begin{table*}[t]
\hspace{0cm}
\caption{Atomic data for Fe II, Fe III and Fe VII.}
\label{1a}
 \begin{minipage}{240mm}
\rotatebox{0}{
  \begin{tabular}{@{}lllll@{}} \hline\hline
Ion    & Levels & Temperature (($\times$1000K)        & A value ref.                & Coll. strength ref. \\ \hline
Fe II  & 46     & 1-20                        & Nussbaumer \& Storey (1988) & Bautista \& Pradhan (1996) \\
       &        &                             & Garstang (1962)             & Bautista, private communication \\  \hline
Fe III & 27     & 3-70                        & Nahar \& Pradhan (1996)     & Zhang (1996)\\ \hline
Fe VII & 9      & 2-1000                      & Berrington et al. (2000); Young et al. (2005) & Berrington et al. (2000) \\ \hline
\label{STIS}
\end{tabular}}
\end{minipage}
\end{table*}
\end{center}

The high resolution VLT spectra were used
to check if any of the lines observed in the much lower resolution AAT spectra suffered from contamination
by another line. For some of the lines ({\it e.g.}, [O II] $\lambda\lambda$7319,7331 and [Ca II] $\lambda$7324) the
spectral resolution of the AAT narrow-slit observations was sufficient to allow deblending the lines using Gaussian fitting.
However, the He I $\lambda$3889 and [Ne III] $\lambda$3968 lines were found to be strongly blended with nearby
lines of hydrogen and were therefore excluded from the analysis. The measured AAT and VLT optical line fluxes at 1416, 1680, 2864, 4309,
5040 days are listed for H, He, N, O, Ne, S, Ar, Ca and Fe in Table 4. The H$\alpha$ and H$\beta$ fluxes at 1416, 1680, 2864, 4309 days were 
corrected for contamination by Star 3. The fluxes measured directly from the wide slit spectra were found to have uncertainties of $\pm$15 - 21\%,
the fluxes measured from the narrow slit spectra uncertainties of $\pm$43\%, and the fluxes measured
from the VLT spectra uncertainties of $\pm$20\% except for [S III] $\lambda\lambda$9069,9532 in the red end of the spectrum
for which $\pm$50\% was adopted. The correction of the H$\alpha$ and H$\beta$ fluxes from 
epochs earlier than 1400 days was not possible without knowing the details of the observations and the 
subsequent data reductions. However, Star 3 contamination at these early epochs is not significant, since the ring was much brighter than
Star 3. To investigate the effects of the collision between the SN ejecta and the inner ring material we also measured the
fluxes for [O III] $\lambda\lambda$4959,5007, [Ne III] $\lambda$3869, [Ne IV] $\lambda\lambda$4714,4726, [Ar III] $\lambda\lambda$7136,7751,
[Ar IV] $\lambda\lambda$4711,4740 and [Fe VII] $\lambda$5721 at 5703, 6618, 6826, 7183, and 7559 days from the VLT spectra. These
fluxes are listed in Table 4.

\section{Modeling the Line Light Curves}
The latest version of the photoionization code used and described in 
Lundqvist $\&$ Fransson (1991), Lundqvist (1992, 1999, 2007) and LF96 was used
here to model the emission line fluxes from the inner ring around
SN 1987A. In the model the inner ring is initially ionized by
the soft X-ray and UV photons emitted in the SN shock break-out. We have used the 500full1 model of Ensman \& Burrows (1992) for the ionization. 
It was confirmed by Blinnikov et al. (2000) to give a good representation of the break-out burst (cf. Lundqvist 2007). Initially
the ring material was assumed to be optically thick to the ionizing flash and therefore, the outer radius
of the ionized gas determined by the number of ionizing photons and the density 
of the gas, {\it i.e.}, the emission region in the ring was assumed to be ionization bounded.
The outer radii of the different model density components were checked and if needed truncated to match
the observed dimensions of the inner ring (see Sect. 3.1).
The SN flash sets up the initial ionization structure of the ring, and the gas
then recombines and cools. The model includes
cooling by ions of H, He, C, N and O with all the ionization stages,
and Ne, Na, Mg, Al, Si, S, Ca, Ar and Fe with the first 9, 10, 11, 12, 11, 11, 11, 11 and 16 ionization stages, respectively. The code 
follows the ionization, recombination and cooling of the gas with time including
collisional and Auger ionization ({\it i.e.}, a high energy photon ejecting an inner shell electron
followed by ejection of an outer shell electron), radiative and dielectric recombination, collisional excitation 
and de-excitation, and charge transfer. As in LF96, the geometry of the ring is modeled as a section of a spherical shell 
with the inner and outer edges at the same opening angle ($\theta$ = 2.5$^{\circ}$) seen from the SN located in the middle of
the circular ring. The ring is divided into 70 concentric ``shells'', the outer ones being more massive, ({\it i.e.},
bigger volume) than the inner ones. This is because the thermal and ionization structure changes more rapidly 
in the inner parts of the ring thus requiring better ``sampling''. 

The input parameters of the code are the mass and radius of the ring, the gas number density, 
the spectral and temporal characteristics of the ionizing SN flash, and the elemental 
abundances in the ring. The program extracts emissivities for the most common UV-IR 
emission lines of H, He, C, N, O, Ne, Na, Mg, Al, Si, S, Ar, Ca and Fe at several ionization 
stages. For many of the ions, multilevel model atoms are used to calculate the emission. While the levels
of ionization are treated time-dependently, the levels within the model atoms are in steady state, except for H and He
where the population of all atomic states are treated time-dependently, with a total of 56 levels for H I-II and 72 levels for
He I-III. Strong resonance lines of all ions, and selected lines between excited states, are treated with escape probabilities, 
taking the limited thickness of the ring into account. In the multilevel atoms of H and He, photoionization is assumed to occur 
only from the ground state. While this is a good approximation to calculate the degree of ionization, the population of some 
excited states with low de-excitation rates could be less well modeled.

Although the ionization fractions for the 16 first ionization stages of iron are calculated
in the photoionization code, the code does not calculate individual line emissivities for many of the iron
ions. Instead, the cooling was pre-calculated using multilevel atoms and tabulated for various densities and
temperatures as described in LF96. The atomic data included in the code are not sufficiently accurate to calculate the populations 
of individual energy levels with an accuracy allowing the iron line emissivities to be determined. Therefore, we stored temperature, 
electron density and levels of ionization for iron at several time steps and ran a separate code to determine the line fluxes. For this 
we collected atomic data for Fe II, Fe III and Fe VII from the literature (see Table 5), to allow accurate modeling of the [Fe II], 
[Fe III] and [Fe VII] line light curves. Thanks to the Iron Project (Hummer et al. 1993), these atomic data have improved significantly 
and make such abundance determinations more reliable. For Fe II, a 46-level model atom was used. We took the Einstein A-values for 
IR transitions from Nussbaumer \& Storey (1988) and the other transitions from Garstang (1962). The more recent A-values by Quinet
{\it et al.} (1996) were not used since they have been observed to yield discrepancies when used in modeling (Bautista \& Pradhan 1998).
 The collision strengths were taken from Bautista \& Pradhan (1996), and Bautista, private communication. As the temperature varies 
strongly with time and across the ring, 8 different sets of collision strengths are used here for temperatures ranging between 1000 
and 20,000 K. For Fe III, a 27-level model atom was used. The Einstein A values were taken from Nahar \& Pradhan (1996), and the 
collision strengths from Zhang (1996). Here 16 different sets of collision strengths were used for temperatures ranging from 3000 
to 70,000 K. For Fe VII we used a 9-level model adopting data from Berrington et al. (2000) and the corrected A-values by Young et al. (2005). 

Finally, equipped with the level populations of the ions of interest the total line fluxes observed from the whole ring were obtained by 
integrating the line emissivities over the volume of the ring (see Lundqvist \& Fransson 1991 and LF96 for details). This takes into 
account the time delay caused by the physical size of the ring and the finite velocity of light.

\begin{figure*}
\centering
\vspace{+0.5cm}
\includegraphics[height=15cm]{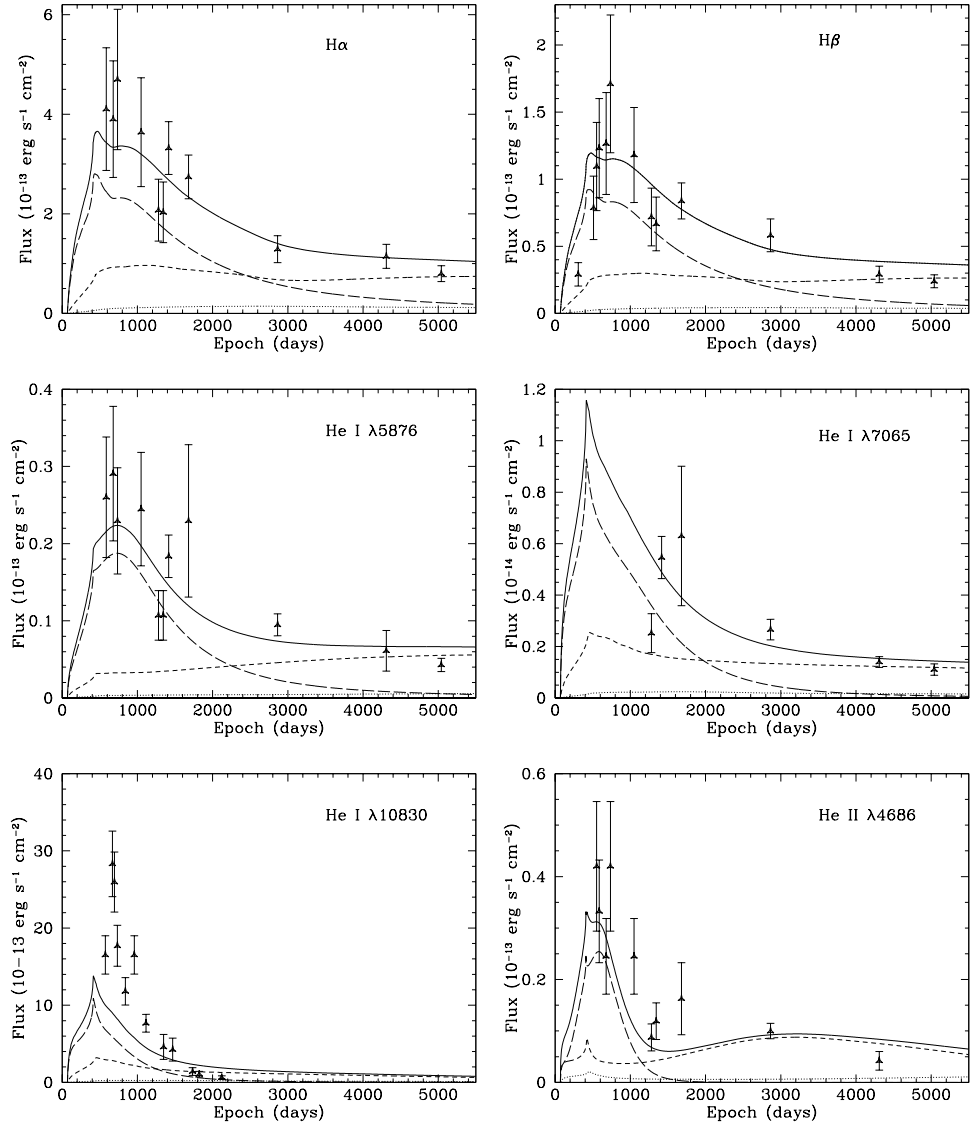}
\vspace{+0cm}
\caption{The observed fluxes of lines of H and He plotted together with the best fit (see Table 6)
model fluxes. The observed line fluxes have been dereddened assuming
E$_{B-V}$ of 0.05 and 0.15 for the Galaxy and the LMC, respectively. 
The long dashed, short dashed and dotted lines correspond to the density components
of 3$\times$10$^{4}$, 3$\times$10$^{3}$, and 1$\times$10$^{3}$ atoms cm$^{-3}$, respectively.
The sum of these models is shown as a solid line.}
\label{fig:1}
\vspace{+0.5cm}       
\end{figure*}

\begin{figure*}
\centering
\vspace{+0.5cm}
\includegraphics[height=15cm]{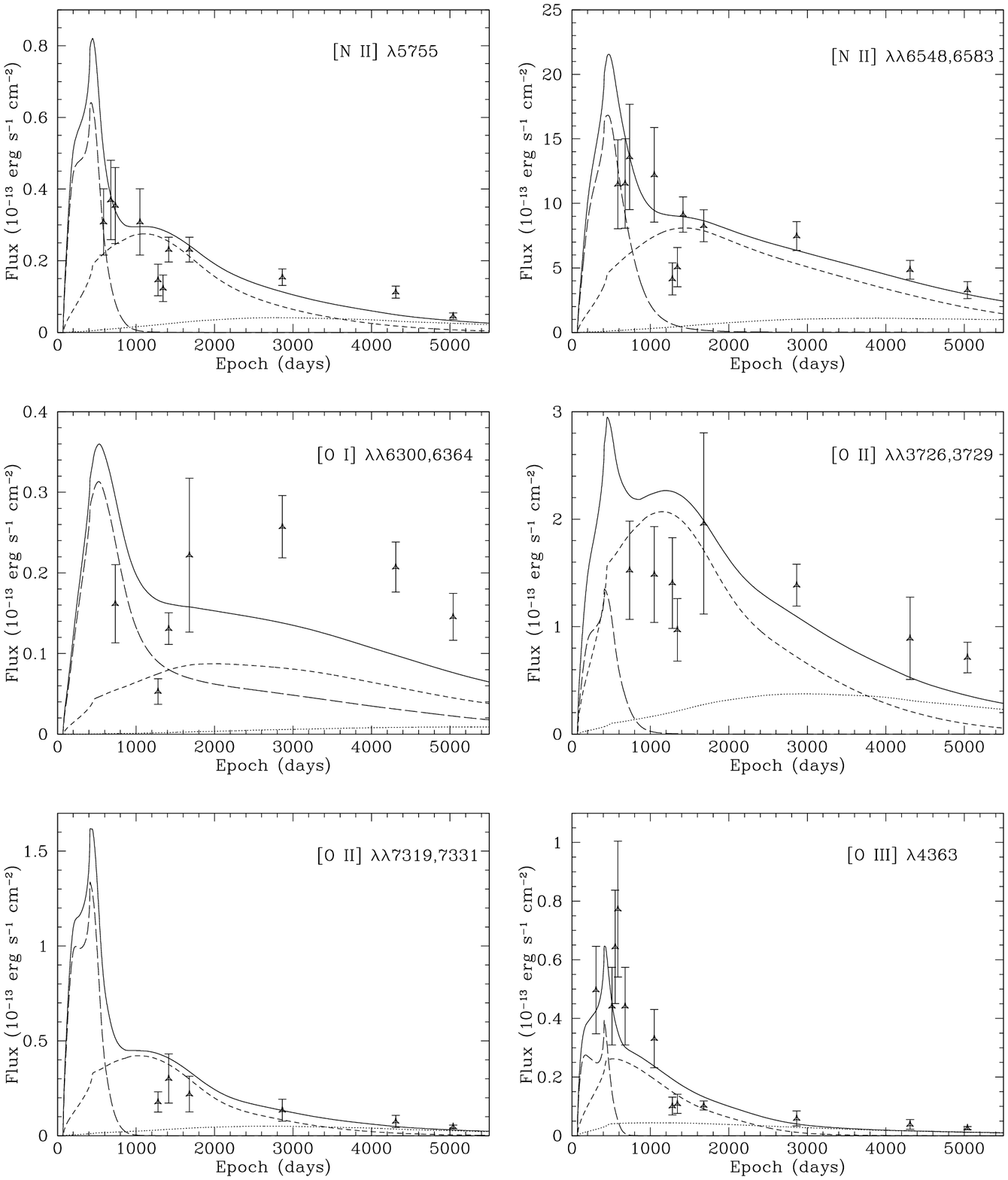}
\vspace{-0cm}
\caption{The observed fluxes of [N II], [O I], [O II] and [O III] lines plotted together with 
the best fit (see Table 6) model fluxes. The observed line fluxes have been dereddened assuming
E$_{B-V}$ of 0.05 and 0.15 for the Galaxy and the LMC, respectively. 
The long dashed, short dashed and dotted lines correspond to the density components
of 3$\times$10$^{4}$, 3$\times$10$^{3}$, and 1$\times$10$^{3}$ atoms cm$^{-3}$, respectively.
The sum of these models is shown as a solid line.}
\label{fig:1}
\vspace{+0.5cm} 
\end{figure*}

\begin{figure*}
\centering
\vspace{+0.5cm}
\includegraphics[height=15cm]{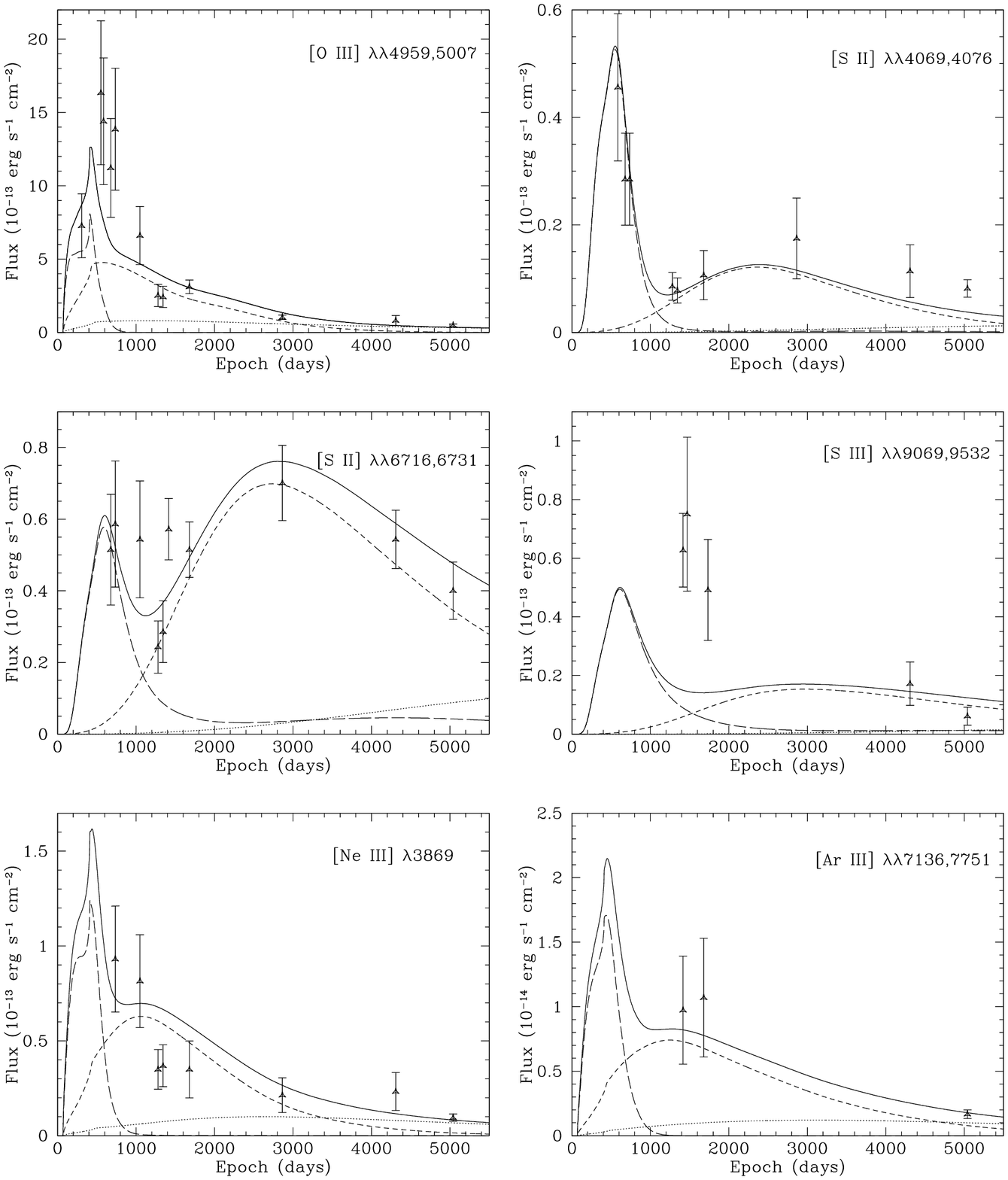}
\vspace{-0cm}
\caption{The observed fluxes of [O III], [S II], [S III], [Ne III] and Ar [III] lines 
plotted together with the best fit (see Table 6) model fluxes. The observed line fluxes have been dereddened assuming
E$_{B-V}$ of 0.05 and 0.15 for the Galaxy and the LMC, respectively. 
The long dashed, short dashed and dotted lines correspond to the density components
of 3$\times$10$^{4}$, 3$\times$10$^{3}$, and 1$\times$10$^{3}$ atoms cm$^{-3}$, respectively.
The sum of these models is shown as a solid line.}
\label{fig:1} 
\vspace{+0.5cm}
\end{figure*}

\begin{figure*}
\centering
\vspace{+0.5cm}
\includegraphics[height=15cm]{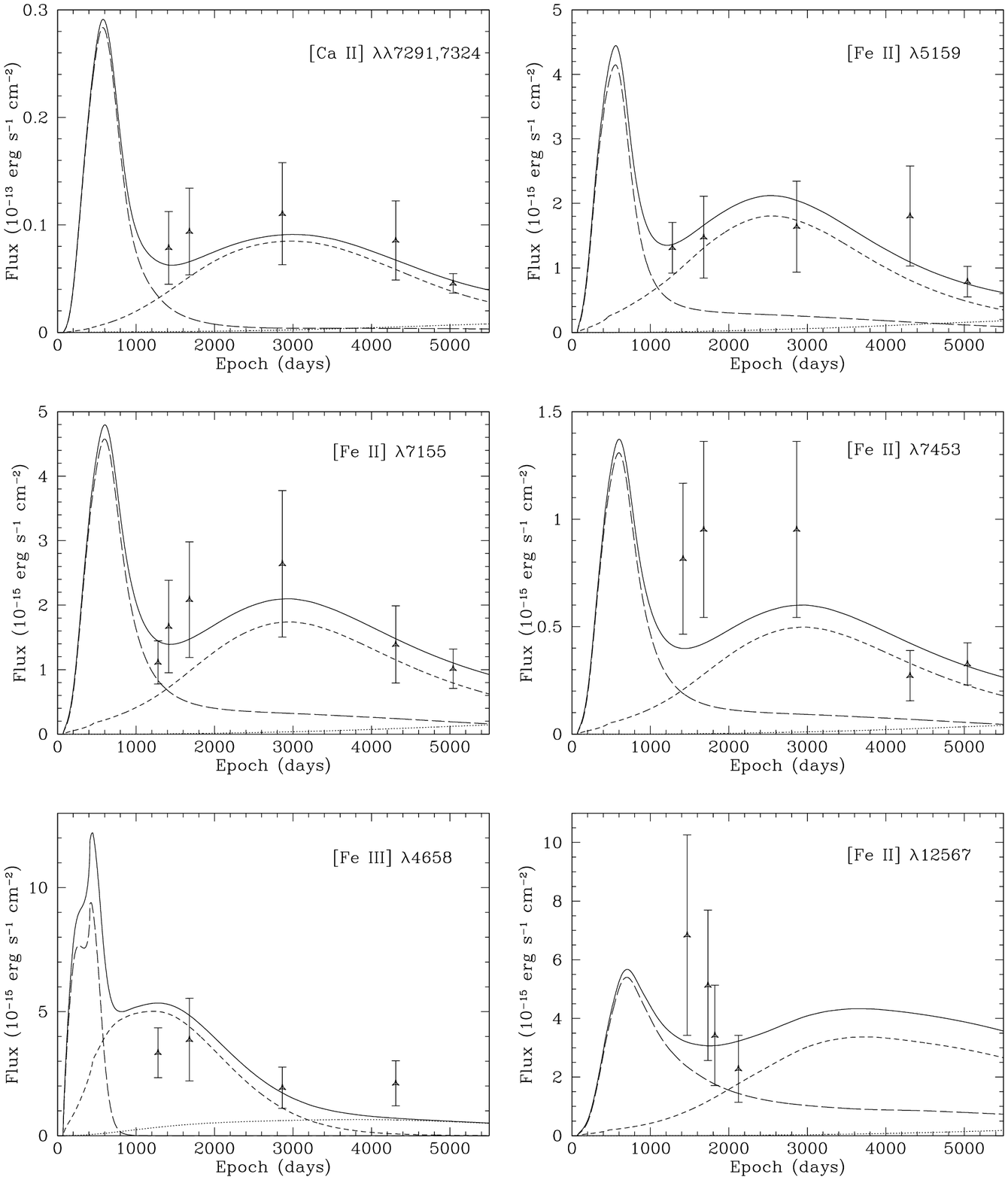}
\vspace{-0cm}
\caption{The observed fluxes of [Ca II], [Fe II] and [Fe III] lines plotted together 
with the best fit (see Table 6) model fluxes. The observed line fluxes have been dereddened assuming
E$_{B-V}$ of 0.05 and 0.15 for the Galaxy and the LMC, respectively. 
The long dashed, short dashed and dotted lines correspond to the density components
of 3$\times$10$^{4}$, 3$\times$10$^{3}$, and 1$\times$10$^{3}$ atoms cm$^{-3}$, respectively.
The sum of these models is shown as a solid line.}
\label{fig:1}   
\vspace{+0.5cm}    
\end{figure*}

\begin{figure*}
\centering
\vspace{+0.5cm}
\includegraphics[height=15cm]{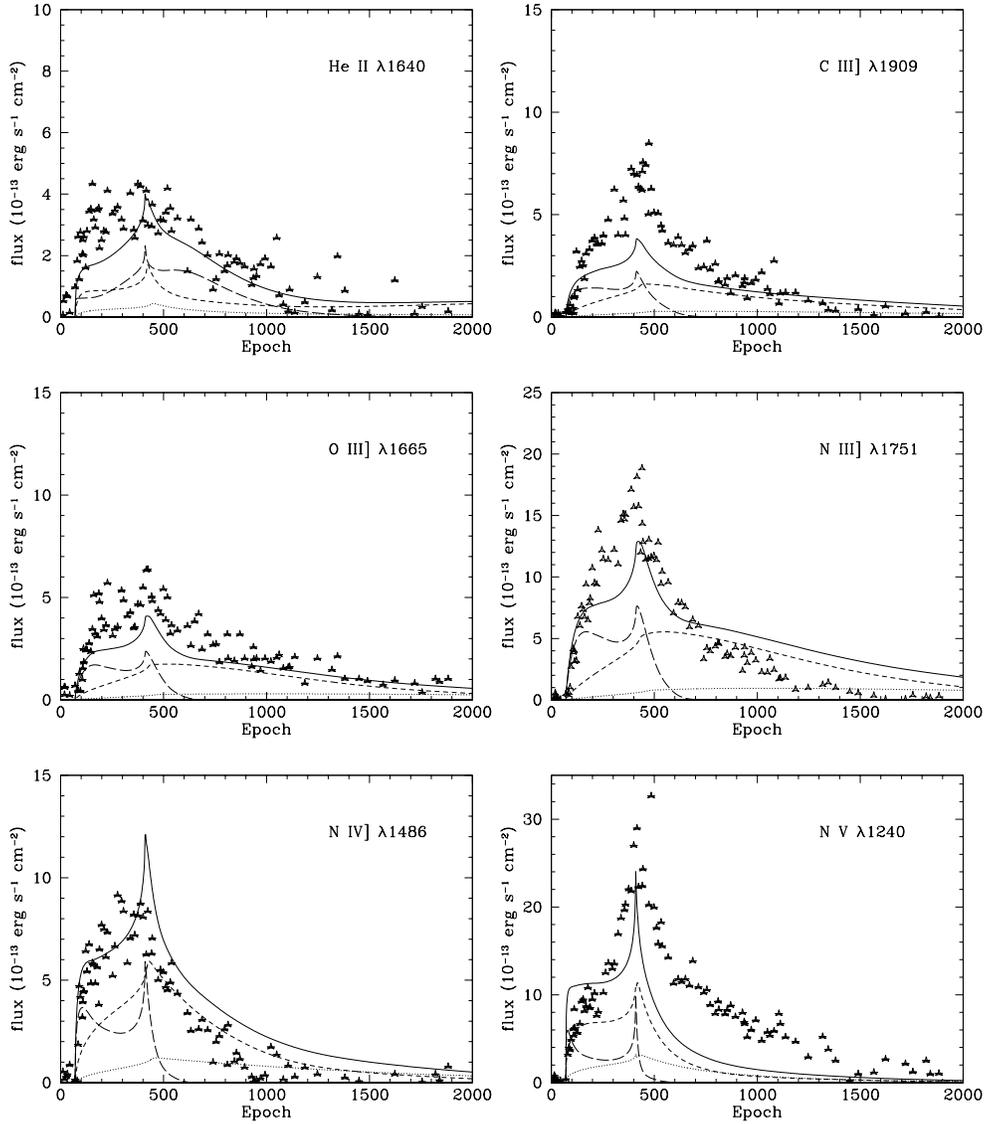}
\vspace{-0cm}
\caption{
The observed fluxes of the He II $\lambda$1640, C III] $\lambda$1909, O III] $\lambda$1665, N III] $\lambda$1751,
N IV] $\lambda$1486, and N V $\lambda$1240 UV lines plotted together with the model fluxes. The observed line fluxes
(from Sonneborn et al. 1997) have been dereddened assuming E$_{B-V}$ of 0.05 and 0.15 for the Galaxy and 
the LMC, respectively. The long dashed, short dashed and dotted lines correspond to the density components
of 3$\times$10$^{4}$, 3$\times$10$^{3}$, and 1$\times$10$^{3}$ atoms cm$^{-3}$, respectively,
using the abundances found in LF96 (see Table 7). The sum of these models is shown as a solid line.}
\label{fig:1}   
\vspace{+0.5cm}    
\end{figure*}

\begin{figure*}
\centering
\vspace{+0.5cm}
\includegraphics[height=15cm]{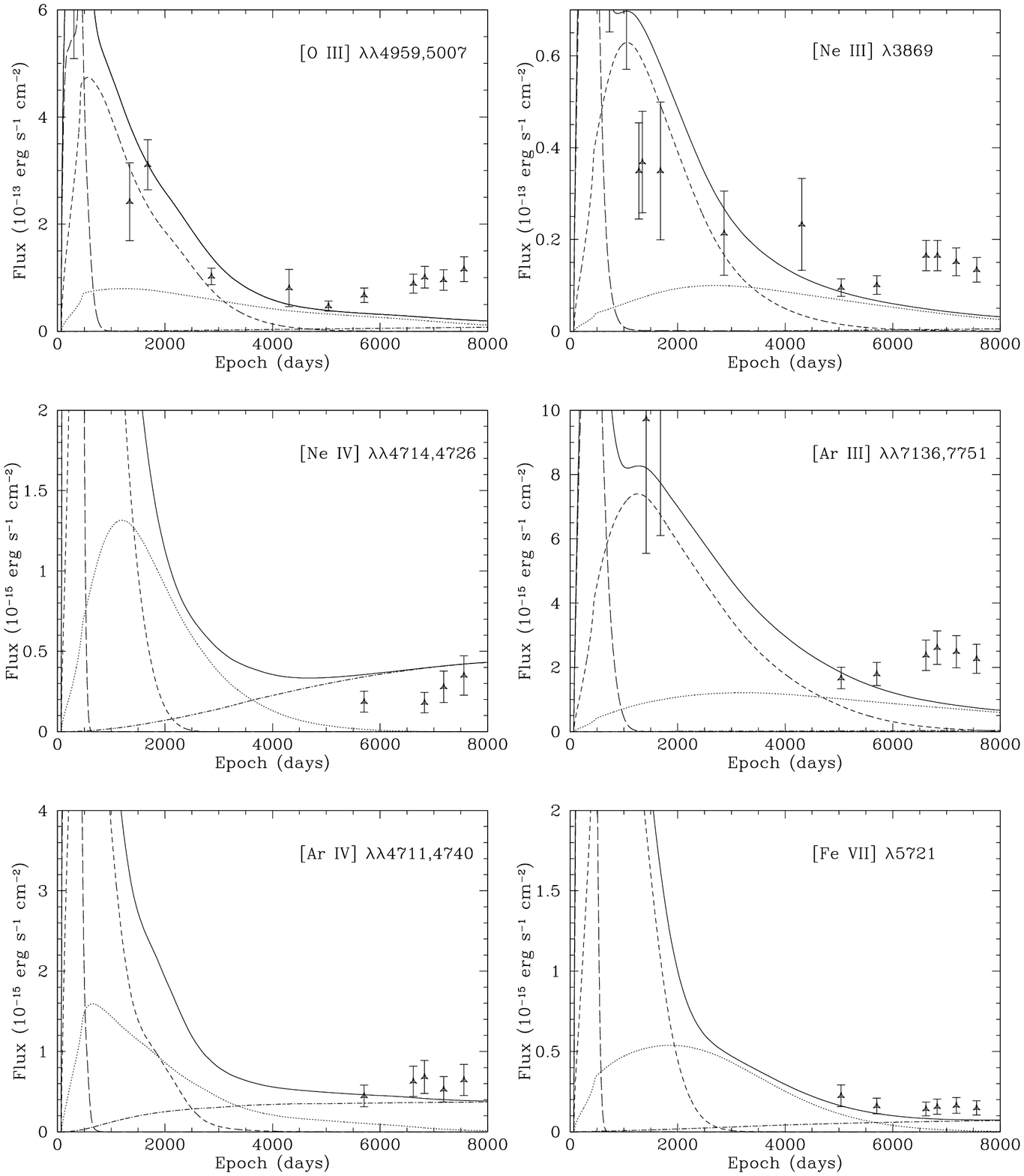}
\vspace{-0cm}
\caption{The observed fluxes of the [O III] $\lambda\lambda$4959,5007, [Ne III] $\lambda\lambda$3869,3968, [Ne IV] $\lambda\lambda$4714,4726,
[Ar III] $\lambda\lambda$7136,7751, [Ar IV] $\lambda\lambda$4711,4740, and [Fe VII] $\lambda$5721 lines plotted together with the model fluxes. 
For oxygen, neon and argon the abundances found when fitting the pre-5000 day light curves were adopted (see Table 7), 
and for iron an abundance higher than the average abundance by one standard deviation was used.
The observed line fluxes have been dereddened assuming E$_{B-V}$ of 0.05 and 0.15 for the Galaxy and the LMC, respectively. 
The long dashed, short dashed, dotted and long-short dashed lines correspond to the density components
of 3$\times$10$^{4}$, 3$\times$10$^{3}$, 1$\times$10$^{3}$ and 1$\times$10$^{2}$ atoms cm$^{-3}$, respectively.
The sum of these models is shown as a solid line.}
\label{fig:1} 
\vspace{+0.5cm}      
\end{figure*}

\subsection{The Density Structure and Abundances}
The dimensions of the inner ring used for the modeling were obtained from the observations by the HST.
The thickness of the ring has been constrained by HST imaging (Jakobsen et al. 1991; Plait et al. 1995)
to be around 10-15$\%$ of the ring
radius in the plane of the sky. This corresponds to $\sim$20$\%$ of the ring radius for a circular ring.
The ionized gas of the 1$\times$10$^{3}$ cm$^{-3}$ density component extends out to 1.26 $\times$ the ring inner radius, which we have assumed 
to be $6.2\times10^{17}$~cm (cf. Lundqvist 1999), and therefore the outer radius of this density component was truncated at $7.4\times10^{17}$~cm 
to match the observed dimensions of the inner ring. The elemental abundances (by number of atoms) used for the modeling here were initially as 
follows: He/H = 0.25, N/C = 4.5, and N/O = 1.0 
with an overall metallicity, Z, of 0.35 times solar, {\it i.e.} Z/(H+He+Z) = 0.35 times the solar ratio. The observed 
line fluxes were corrected for extinction using the same extinction laws for the LMC and the Galaxy as in LF96, 
assuming an E$_{B-V}$ of 0.05 and 0.15 for Galaxy and the LMC, respectively. 
The extinction laws from Fitzpatrick (1985), 
and Savage $\&$ Mathis (1979) were used for the LMC and the Galaxy, respectively. The dereddened line fluxes
were initially compared with the models by eye. The mass and density structure of the ionized gas in the inner 
ring was determined by searching for satisfactory model fits for (1) the absolute line 
fluxes of H$\alpha$ and H$\beta$, and (2) the observed time evolution of the line fluxes of all the other 
elements. Guided by the previous studies (LF96), 3$\times$10$^{4}$ atoms cm$^{-3}$ was selected as the highest
density component. Lower density components were also discussed in LF96, and the lowest density determined so far is 
that of Mattila (2002), Mattila et al. (2003) and Gr\"oningsson et al. (2008b) who find indications for densities down to 10$^{3}$
atoms cm$^{-3}$. We indeed find that at least two additional components, 1$\times$10$^{3}$ and 3-5$\times$10$^{3}$ 
atoms cm$^{-3}$, were required, in addition to that of  3$\times$10$^{4}$ atoms cm$^{-3}$, to produce the observed line light curves.

To decide on the weights ({\it i.e.}, the masses of ionized gas) for each density component we sought for the optimal
set of weights to give the lowest $\chi$$^{2}$ value for the model fit simultaneously to H$\alpha$, $[$N II$]$ $\lambda\lambda$6548,6583, 
[O III] $\lambda\lambda$4959,5007 and [S II] $\lambda\lambda$6716,6731. The line fluxes at 1280 and 1344 days appear to be
systematically lower than the fluxes from the other authors at similar epochs and therefore we decided not to include
these points in the $\chi$$^{2}$ fits. Also, we were not able to find a satisfactory fit for the H$\beta$ line fluxes
at 307 days and decided to concentrate on the epochs later than this for all the lines. We found that the optimal fit was 
obtained using 0.65, 1.30 and 0.45 as the weights for the 1 $\times$ 10$^{3}$, 3 $\times$ 10$^{3}$ and 3 $\times$ 10$^{4}$ atoms cm$^{-3}$ density 
components, respectively. A slightly higher $\chi$$^{2}$ value was found for the 1 $\times$ 10$^{3}$, 5 $\times$ 10$^{3}$ and 3 $\times$ 10$^{4}$ atoms 
cm$^{-3}$ combination of densities. The masses of ionized gas in the three density components were about 1.2, 3.4, and 1.2 $\times$ 10$^{-2}$ 
M$_{\odot}$, respectively. The lowest-density components dominate the line light curves at the later times. High resolution 
HST images indicate that this low-density gas is situated, on average, further away from the SN than the gas with a
higher density (Lundqvist \& Sonneborn 1997). However, in reality there is probably a continuous 
range of densities within the ring rather than a few discrete density components. 

The elemental abundances yielding the observed absolute line fluxes were then 
determined. This was done by adjusting the total number of emitting 
ions relative to hydrogen. The effects of the modified abundances on the
temperature and ionization structure of the ring were also included
by rerunning the photoionization code after initially finding the required
abundances. The optimal abundances were then determined by searching for the number of 
emitting ions yielding the lowest $\chi$$^{2}$ value for each line (see Table 6).
The resulting model fits are shown in Figs. 3-6 with the contributions of the
different density components indicated.

\subsection{UV Lines and Average Abundances}
The UV emission line fluxes from the inner ring were modeled by LF96 up to $\sim$2000 days from the explosion.
They found gas densities of 6 $\times$ 10$^{3}$, 1.4 $\times$ 10$^{4}$, and 3.3 $\times$ 10$^{4}$ atoms cm$^{-3}$,
and derived abundances for He, C, N and O (see Table 7). In Fig. 7 we compare the model fluxes with the
observations of a few UV lines to test how well our selected model parameters can explain the UV data.
For these plots the same density components were used as for the optical lines whereas
the abundances were adopted from LF96. We find that for the UV lines of helium and oxygen our model
parameters and the abundance estimated in LF96 give reasonable fits, especially at the later epochs,
giving more confidence in our abundance estimates for these elements. However, to satisfactorily model the UV lines of 
carbon and nitrogen a different combination of densities and abundances would be required. For both the elements
the line light curves were successfully modeled in LF96 and in Lundqvist et al. (2007). However, for modelling the
N V $\lambda$1240 line light curve LF96 found that resonance scattering in the circumstellar gas external to the
inner ring was required. This is not included in the models used for this study. We note that the different abundances derived
in LF96 and here should provide a useful estimate of the
uncertainty in the abundances. We therefore decided to also include the abundances based on all the UV
lines studied in LF96 except N V $\lambda$1240 when estimating the average abundances with their associated uncertainties.
Therefore these abundances are also included in Table 6 together with the estimates based on the optical 
lines.

Based on the data in Table 6, the average abundances and their standard deviations were estimated for each element
and these are listed in Table 7. However, we decided to exclude a few lines when calculating the average abundances.
These were He~I $\lambda$10830, [O~I] $\lambda\lambda$6300,6364, and the near-IR [Fe~II] line at 1.6435$\mu$m. 
The He~I $\lambda$10830 data yielded He/H = 0.54 which is $\sim$3 times higher 
than obtained from the other He~I lines.
We note that a similar very high helium abundance of $\sim$0.45 was also obtained by Allen,
Meikle $\&$ Spyromilio (1989) from their narrow line He I $\lambda$10830 observations at 400-700 days.
Fig. 3 shows that He~I $\lambda$10830 flux is underproduced in our models at epochs earlier than 1500 days, but
overproduced later being more consistent with the abundances derived from the optical He I lines. Including
only the three last points in the fitting yields He/H = 0.23 which is similar to the estimates obtained
from the other lines. This could point at problems with fluxing in the earlier He~I $\lambda$10830 data. It is, however, 
well-known that He~I $\lambda$10830 is notoriously difficult to model accurately because of the highly metastable 2~$^3S$ level which allows for optical
depth effects to become important for transitions to this level (e.g., Osterbrock $\&$ Ferland 2008). We have included this 
in our models and are able to boost the He~I $\lambda$10830 / He~I $\lambda$5876 ratio by $10-20$\% compared to the numbers in 
Porter et al. (2005). The line in our sample which is most attenuated by optical depth effects is He~I $\lambda$7065, and the fact that 
our models give the same helium abundance for the two He~I $\lambda\lambda$5876, 7065 lines suggests that optical depth effects are included adequately.
We caution that photoionization from excited states (cf. above), is not included in our models, but it is not clear how that could boost He~I $\lambda$10830 
and not other transitions to the 2~$^3S$ level. We, however, leave the more detailed investigation of the
He~I $\lambda$10830 line light curve for a future study and do not include it here in our estimate for the helium abundance.
The near-IR [Fe~II] line at 1.6435$\mu$m is also excluded due to fluxing difficulties at these wavelengths (see Fassia et al. 2002).

For the [O~I] $\lambda\lambda$6300,6364 line, we found that our combination of densities could not be used to obtain 
a satisfactory model fit. Furthermore, our fit to this line yielded an oxygen
abundance $\sim$4 times (or 15$\sigma$) higher than the average abundance found using five lines of
O~II and O~III.  Assuming that the ring contains high-density regions that are surrounded
by low-density gas, some of this low-density gas will be shielded from the most intense radiation from
the SN flash and will therefore contain more neutral, or at least less ionized gas than if no shielding
was taking place. In fact, since mid-1990 HST images have shown many "hot spots" appearing in different
locations around the entire inner ring ({\it e.g.}, Lawrence et al. 2000; Sugerman et al. 2002) as a result
of the SN ejecta impacting "fingers" of dense ring material protruding towards the SN (see Pun et al. 2002).
In our model shielding by such higher density regions is not included and all the density components see the
SN flash the same way. We expect the higher ionization lines of oxygen to be less affected by the effects of shielding
and therefore yield better model fits and more reliable abundance estimates in this study. The shielding effect
should also affect other low-ionization lines (cf. below).

\subsection{Late Epochs}
As discussed in Lundqvist (1999, see also Chevalier \& Dwarkadas 1995), gas in the H~II-region interior to the inner ring, 
but probably also extending to larger radii than the ring at higher latitudes, can contribute to the observed line emission 
at late epochs. The density of this gas is most likely not higher than $\sim 1.5\times10^{2}$ atoms cm$^{-3}$. Such low 
densities can retain an appreciable fraction of highly ionized gas like N~V even at epochs as late as 6000 days. The main 
uncertainty is how much of the H~II-region has been swept up by the SN blast wave at late epochs. From Fig. 3 of 
Lundqvist (1999) we note that after 4000-5000 days this effect can introduce an uncertainty larger than a factor of two for 
the modeled line emission from the H~II-region. As opposed to the inner ring, the H~II-region was already partially ionized
prior to the SN explosion by the progenitor star. It therefore will not be heated as efficiently by the supernova flash as the inner ring. 
Nevertheless, we have chosen to include a $10^{2}$ atoms cm$^{-3}$ component in our models in the same way as the higher density
components {\it i.e.} we have not included the effects of the earlier ionization, which only affects ions with the ionization
potential of H~I, or lower. The supernova can ionize gas with a $10^{2}$ atoms cm$^{-3}$ density out to 2.23 $\times$ the ring inner
radius. We, truncated the outer radius of this component in the same fashion as for the 10$^{3}$ atoms cm$^{-3}$ component (see Sect. 3.1)
but to correspond to 30\% of the ring inner radius, to agree with the discussion in Lundqvist \& Sonneborn (1997). We compared the
post-5000 day light curves with the models adding the $10^{2}$ atoms cm$^{-3}$ component to the combination of density 
components found in Sect. 3.1 (see Fig. 8). For [O~III], [Ne~III], [Ne~IV], [Ar~III] and [Ar~IV] the average abundances found when fitting the
pre-5000 day light curves were adopted (see Table 7). However, for the [Fe VII] $\lambda$5721 line light curve an iron
abundance higher than the average by one standard deviation was required to satisfactorily model the data.
After the truncation the $10^{2}$ atoms cm$^{-3}$ component was multiplied by 6 (corresponding to a mass of
ionized gas of 1.8 $\times$ 10$^{-2}$ M$_{\odot}$ to yield satisfactory fits to
the observed line fluxes at $\sim$5000 days. While the contribution of the $10^{2}$ atoms cm$^{-3}$ component 
was negligible for the [O III], [Ne III] and [Ar III] lines, it was required to satisfactorily model the [Ne~IV], [Ar~IV] 
and [Fe~VII] line light curves. Assuming a constant density between the inner ring radius and out to $1.3\times$ this radius,
the low-density gas would fill a region with an opening angle of 7.2$^{\circ}$ above and below the ring plane in addition to
the gas in the ring.

At epochs later than $\sim 5000$ days a strong contributor to the narrow line emission is the re-ionization of the ring by soft X-rays 
from the hydrodynamic interaction of the ring with the supernova ejecta. Gr\"oningsson et al. (2008b) demonstrate clearly how 
especially [O~III] and [Ne~III] lines increase in flux between $\sim 5000$ and 6800 days after the explosion due to this effect. 
Later, at least some of these lines appear to level off, or decrease in flux. We have not attempted to model the hydrodynamic interaction. 
However, we do, as suggested in Lundqvist (1999), from day 1150 include modest ionization to simulate the early part of the light 
curve of soft X-rays (Hasinger et al. 1996). We have extrapolated the linear increase of X-ray emission assumed in Lundqvist (1999) 
to late times. As shown in Park et al. (2007), the soft X-ray emission starts to increase much faster after $\sim 4700$ days than 
we have used in our modeling. We have refrained from trying to fine tune our models with detailed fits to the X-ray emission. This 
means that our models should underproduce the emission of at least some lines after $\sim 5000$ days. The amount of underproduction 
is an indication of how much emission is the result of a shock precursor of the ejecta/blob collisions. In Fig. 8 this effect can
be clearly seen for most of the lines considered here.

\begin{table}
\hspace{0cm}
\caption{Abundances for different lines}
\label{1a}
  \begin{tabular}{lccc} \hline\hline
line                              & $\chi$$^{2}$ & N/N$_{H}$ & Use for abundance \\ \hline
H$_{\alpha}$                      & 0.9       & 1.02 & ... \\  
H$_{\beta}$                       & 1.3       & 0.87 & ... \\ \hline
He I $\lambda$5876                & 1.8       & 0.175 & Yes \\
He I $\lambda$7065                & 1.5       & 0.159 & Yes \\
He I $\lambda$10830               & 13        & 0.54  & No \\
He II $\lambda$1640               & ...       & 0.25  & Yes \\
He II $\lambda$4686               & 2.0       & 9.23$\times$10$^{-2}$ & Yes \\ \hline
C III$]$ $\lambda$1909            & ...       & 3.24$\times$10$^{-5}$ & Yes \\ \hline
$[$N II$]$ $\lambda$5755          & 3.2       & 3.54$\times$10$^{-4}$ & Yes \\
$[$N II$]$ $\lambda\lambda$6548,6583     & 1.0       & 3.95$\times$10$^{-4}$ & Yes \\
N III$]$ $\lambda$1751            & ...       & 1.82$\times$10$^{-4}$ & Yes \\ 
N IV$]$ $\lambda$1486             & ...       & 1.82$\times$10$^{-4}$ & Yes \\ \hline
$[$O I$]$ $\lambda\lambda$6300,6364      & 7.6       & 7.88$\times$10$^{-4}$ & No \\
$[$O II$]$ $\lambda\lambda$3726,3729     & 2.9       & 1.76$\times$10$^{-4}$ & Yes \\
$[$O II$]$ $\lambda\lambda$7319,7331     & 1.5       & 2.51$\times$10$^{-4}$ & Yes \\
O III$]$ $\lambda$1665            & ...       & 1.58$\times$10$^{-4}$ & Yes \\
$[$O III$]$ $\lambda$4363         & 2.7       & 1.57$\times$10$^{-4}$ & Yes \\
$[$O III$]$ $\lambda\lambda$4959,5007    & 2.7       & 1.88$\times$10$^{-4}$ & Yes \\ \hline
$[$Ne III$]$ $\lambda$3869        & 1.0       & 8.59$\times$10$^{-5}$ & Yes \\ \hline
$[$S II$]$ $\lambda\lambda$4069,4076     & 1.9       & 1.72$\times$10$^{-5}$ & Yes \\
$[$S II$]$ $\lambda\lambda$6716,6731     & 1.1       & 1.76$\times$10$^{-5}$ & Yes \\
$[$S III$]$ $\lambda\lambda$9069,9532    & 7.0       & 5.04$\times$10$^{-6}$ & Yes \\ \hline
$[$Ar III$]$ $\lambda\lambda$7136,7751   & 0.3       & 1.68$\times$10$^{-6}$ & Yes \\ \hline
$[$Ca II$]$ $\lambda\lambda$7291,7324    & 0.4       & 3.22$\times$10$^{-6}$ & Yes \\ \hline
$[$Fe II$]$ $\lambda$5159         & 0.4       & 1.78$\times$10$^{-5}$ & Yes \\
$[$Fe II$]$ $\lambda$7155         & 0.3       & 8.19$\times$10$^{-6}$ & Yes \\
$[$Fe II$]$ $\lambda$7453         & 1.4       & 7.63$\times$10$^{-6}$ & Yes \\ 
$[$Fe II$]$ $\lambda$1.2567$\mu$m & 0.8       & 3.67$\times$10$^{-6}$ & Yes \\
$[$Fe II$]$ $\lambda$1.6435$\mu$m & 1.5       & 3.67$\times$10$^{-6}$ & No \\
$[$Fe III$]$ $\lambda$4658        & 1.4       & 1.03$\times$10$^{-5}$ & Yes\\ \hline
\end{tabular}
\\
Best fit reduced $\chi$$^{2}$ values and the resulting abundances (relative to hydrogen) for different lines
(see Figs. 3-6). Whether the line is included in the average abundance (see Table 8) is indicated in
col. 4.
\vspace{+0.5cm}
\end{table}

\begin{table*}
\hspace{0cm}
\caption{Elemental abundances}
\label{1a}
 \begin{minipage}{240mm}
  \begin{tabular}{ccccccccccc} \hline\hline
Element  & solar(AG89)   & solar(GAS07)  & LMC(+)      & LMC(X-ray,H) & LMC(H07)     & 87A(X-ray,D) & 87A(X-ray,Z) & 87A(LF96) & This work & Lines \\ 
\hline
  He     & 10.93(0.01,*) & 10.93(0.01,*) & 10.94(0.03) &    ...      & ...         & ...         &  ...        & 11.40     & 11.23$^{+0.13}_{-0.19}$ & 4\\
  C      & 8.56(0.04)    &  8.39(0.05)   &  8.04(0.18) &    ...      & 7.75         & ...         &  ...        & 7.51      & ...        & ...\\
  N      & 8.05(0.04)    &  7.78(0.06)   &  7.14(0.15) &    ...      & 6.90         & 7.81(0.07)   &  7.81(0.06)  & 8.26      & 8.44$^{+0.15}_{-0.22}$  & 4\\
  O      & 8.93(0.04)    & 8.66(0.05)    & 8.35(0.06)  & 8.21(0.07)   & 8.35         & 7.85(0.06)   &  7.85(0.05)  & 8.20      & 8.27$^{+0.08}_{-0.10}$  & 5\\
  Ne     & 8.09(0.10)    & 7.84(0.06)    & 7.61(0.05)  & 7.55(0.08)   & ...         & 7.56(0.10)   &  7.55(0.03)  & ...      & 7.93     & 1\\
  S      & 7.21(0.06)    & 7.14(0.05)    & 6.70(0.09)  & 6.77(0.13)   & ...         & 6.69(0.11)   &  6.68(0.09)  & ...      & 7.12$^{+0.19}_{-0.34}$  & 3\\
  Ar     & 6.56(0.10)    & 6.18(0.08)    &  6.29(0.25) &    ...      & ...         & ...         &  6.29        & ...      & 6.23  & 1\\
  Ca     & 6.36(0.02)    & 6.31(0.04)    & 5.89(0.16)  &    ...      & ...         & ...         &  5.89        & ...      & 6.51  & 1\\
  Fe     & 7.67(0.03)    & 7.45(0.05)    & 7.23(0.14)  & 7.01(0.11)   & ...         & 6.97(0.04)   &  6.97(0.02)  & ...      & 6.98$^{+0.19}_{-0.34}$  & 5\\
\hline
\end{tabular}
\\
The elemental abundances are in units of 12 + log[n(X)/n(H)] and the errors are given in parentheses.
The abundance estimates from\\
this work are the average of the estimates from different emission lines (see Table 6) and the quoted errors 
are the standard deviations.\\ The number of
emission lines used for each average abundance is given in col. 11. (*) = GAS07 = Grevesse, Asplund \& Sauval (2007)\\ (values are for
photospheric abundances). AG89 = Anders \& Grevesse (1989). (+) = Russell \& Dopita (1992). H = Hughes, Hayashi,\\ Koyama (1998).
D = Dewey et al. (2008). H07 = Hunter et al. (2007). Z = Zhekov et al. (2009). LF96 = Lundqvist \& Fransson (1996).\\
\end{minipage}
\vspace{+0.5cm}
\end{table*}

\section{Discussion}
\subsection{Abundances}
The main aim of this study was to determine the elemental abundances of the gas in the inner ring of SN 1987A. The results are 
summarised in Table 7, where we also list abundances found from fits to X-ray grating spectra of the ring collision (Dewey et al. 2008,
Zhekov et al. 2009), as well as for the LMC (Russell \& Dopita 1992, Hughes et al. 1998, Hunter et al. 2007) and for solar elemental abundances (Anders \& Grevesse 1989, Grevesse et al. 2007).
For the inner ring, LF96 derived the abundances 11.40, 7.51, 8.26, 8.20 for He, C, N, O, respectively, in the same units as in Table 7, i.e., $12 + {\rm log}[n({\rm X})/n({\rm H})$] where 
$n({\rm X})$ is the number density of element X. Those numbers were obtained from fits to mainly UV lines observed with $IUE$. In the current paper we find 11.20, 8.41, 8.27, for He, N and O, respectively. No carbon line exists in this optical data set.

The average He/H-ratio found using the three optical helium lines considered in this study is $0.14\pm0.04$, and $0.17\pm0.06$ when including also the He II $\lambda$1640 UV line from LF96.
The large error is due to the He~II~$\lambda 4686$ line yielding $\sim$2$\times$ lower abundance than the rest of these lines. This helium abundance is lower than, but within the 
errors of, the $0.25\pm0.05$ derived by LF96. We believe that our result is more robust than that of LF96 since their abundance was obtained from a single, excited He~II line in the UV where the uncertainty in the reddening correction has a larger effect than in the optical. Hydrogen and helium have also been updated in our models (cf. above) since LF96. He/H = 0.17 is $\sim 2.0$ times larger 
than the solar and the LMC values. Abundances derived from X-ray models of the ring collision (e.g., Zhekov et al. 2009) have used the He and C abundances of LF96 and then varied other 
abundances to make best fits. With He/H = 0.17, instead of the 0.25 in LF96, the X-ray based abundances would need to be somewhat revised.

The nitrogen abundance we find using only the optical [N~II] lines from this study is twice as high as the one found in LF96 whereas the oxygen abundances from these two studies 
are almost identical. The N/O-ratio based on only the optical lines is $1.8\pm0.4$, which is therefore significantly higher than the one found by LF96 of N/O = $1.1\pm0.4$.
Including also the UV lines from LF96 in our abundance estimates the N/O-ratio becomes $1.5\pm0.7$, which is higher than, but still within the errors of, the one found by LF96
and the ratio N/O $\sim$ 0.9 found by the X-ray studies (Dewey et al. 2008, Zhekov et al. 2009). In Lundqvist (2007) we redid the calculations from LF96 with the most recent 
version of the flash-ionization code, and obtained good fits to both the UV line light curves and the optical [N~II]~$\lambda\lambda$6548,6583 and [O~III]~$\lambda\lambda$4959,5007 
line light curves for epochs earlier than 1400 days for the same abundances as in LF96. The chief differences between the LF96 models and those in the current paper are mainly 
due to the difference in density distribution and that we now concentrate on data at later epochs. In LF96 we could not constrain gas with densities lower than $6\times10^3$~cm$^{-3}$, whereas our 
study here is mainly sensitive to lower densities. We note that both the study of LF96 and X-ray studies are likely to probe the gas on the inside of the ring where 
the ring densities probably are higher than further out (cf. the discussion in Lundqvist \& Sonneborn 1997). Here, we have concentrated on the low-density gas and have 
assumed that this is not shielded from the UV flash by high-density gas closer to the supernova. This may not be true for all low-density gas and thus there could be a 
systematic difference between LF96 and the present analysis. Our estimated N/O-ratio is much higher than the range 0.03 $<$ N/O $<$ 1 found for a sample LMC B-type supergiants 
by Hunter et al. (2007). However, a very high N/O-ratio of $\sim$2.1 has also been found for the circumstellar nebula around the Galactic blue supergiant Sher 25
by Hendry et al. (2008). Sher 25 has a spectral type similar to the progenitor of SN 1987A at the time of its explosion. Furthermore, Sher 25 is also surrounded 
by an hourglass shaped circumstellar nebula that is believed to have been ejected by the star about 6600 years ago and has been suggested to have many similarities
to the ring structure of SN 1987A (Brandner et al. 1997). In the case of Sher 25 Hendry et al. found also the photospheric N/O-ratio to be very high and consistent with the one
they derived for the circumstellar nebula.

Adopting the carbon abundance from LF96, together with He/H = $0.17\pm0.06$ and the N and O abundances found using both optical and UV lines in this paper, (N+O)/H = $(4.7\pm1.2)\times10^{-4}$,
the total (C+N+O)/(H+He) abundance is $(4.3\pm1.0)\times10^{-4}$. This can be compared with the old solar value of Anders \& Grevesse (1989), $(1.3\pm0.1)\times10^{-3}$, 
the new solar value of Grevesse et al., $(7.1\pm0.8)\times10^{-4}$, and the LMC value according to Russell \& Dopita (1992), $(3.3\pm0.8)\times10^{-4}$. In the recent paper 
of Hunter et al. (2007), the present-day LMC abundances of (C+N+O)/(H+He) for B-type stars is argued to be $\sim 2.7\times10^{-4}$, i.e., consistent with the findings of Russell 
\& Dopita (1992). Our results are roughly 1.6 times higher than found for the LMC, and about 60\% of the modern solar values.
The explanation could be due to the high helium abundance in SN 1987A. A simple test is to use LMC abundances of metals and helium, and to elevate these compared to hydrogen with 
the same factor as the He/H ratio in the inner ring of SN 1987A, i.e., with a factor of $\sim 2$ to simulate the loss of hydrogen from the surface of the progenitor. Doing 
so, the (C+N+O)/(H+He) abundance would be $\sim 4.2\times10^{-4}$, i.e., nearly the same as we found for the inner ring of SN 1987A.

The iron abundance was estimated based on three [Fe~II] lines and one [Fe~III] line yielding 0.20 $\pm$ 0.11 times the solar value of Anders \& Grevese (1989), which is
consistent with the findings from the X-ray studies (Dewey et al. 2008, Zhekov et al. 2009) of $\sim 0.2 \times$solar. It is also well within the range generally observed in the LMC, which is $\sim (0.2 - 0.3) \times$solar (Russell \& Dopita 1992; Hughes et al. 1998). An iron abundance significantly lower than this could have indicated depletion onto dust grains in 
the ring material. Also, a silicon abundance of the X-ray emitting gas was found by Zhehov et al. (2009) to have a value typical for the LMC. However, recent Spitzer observations (Bouchet et al. 2006;
Dwek et al. 2008) have found evidence for a significant amount of $\sim$10$^{-6}$ M$_{\odot}$ of silicate grains within the circumstellar ring.

Our estimates for the neon, argon and calcium abundances are all based only on a single (doublet in the case of Ar and Ne) emission line and therefore,
these abundances could be considered less reliable. However, for neon and argon we also compared late time (post 5000 day) light curves of
[Ne~IV] $\lambda$4714+4726 and [Ar~IV] $\lambda$4711+4740 lines with the models (see Fig. 8). For this their abundances found earlier (Table 7)
were used and an additional 10$^{2}$ atoms cm$^{-3}$ density component was included (see Sect. 3.2) giving a satisfactory match at $\sim$5000 days
for all the six lines modeled. Also, the evolution of the [Ne~III] doublet is expected to behave similar to the strong and well-sampled [O~III] doublet. 
This gives us more confidence on the estimated neon and argon abundances and we conclude that they are probably accurate to a similar level
as found for iron, {\it i.e.}, $\pm$$\sim$50$\%$. Within these uncertainties we find that the abundances of both argon and neon
are consistent with their LMC values and also with the abundances found in the X-ray studies of the inner ring. However, our abundances
for calcium and sulphur are both higher than the ones found for the LMC and for the ring in the X-ray studies. We note that both the  [S~II] and
[Ca~II] lines used in our study are prone to the shielding effect discussed in Sect. 3.2 for [O~I], and we should thus only consider our abundances for S and Ca
as upper limits.

\subsection{Density Components and Late Epochs}
We found that satisfactory model fits could be produced for the observed line light curves using a combination of at least three
different components with atomic densities of 1 $\times$ 10$^{3}$, $\sim$3 $\times$ 10$^{3}$, 3 $\times$ 10$^{4}$ atoms cm$^{-3}$.
The outer radius of the 1 $\times$ 10$^{3}$ component needed to be truncated to correspond to the observed dimensions of the ring whereas the
higher density components were ionization bounded. The masses of ionized gas in the three density components were about 
1.2, 3.4, and 1.2 $\times$ 10$^{-2}$ M$_{\odot}$, respectively, giving a total mass of ionized gas of $\sim$5.8 $\times$ 10$^{-2}$
M$_{\sun}$. For comparison, LF96 modeled the ring using three components 6 $\times$ 10$^{3}$, 1.4 $\times$ 10$^{4}$, and 3.3 $\times$ 
10$^{4}$ atoms cm$^{-3}$ with masses of 2.4, 1.0, and 1.2 $\times$ 10$^{-2}$ M$_{\odot}$, respectively. Therefore, our total mass of 
ionized gas is $\sim 25\%$ higher than found by LF96, which is not surprising since we are here able to probe a broader range of density 
components than in LF96.

We also compared late time ($\sim$5000 - 7500 days) light curves of [O~III], [Ne~III], 
[Ne~IV],  [Ar~III], [Ar~IV], and [Fe ~II] to the photoionization models and found that an additional $10^{2}$ atoms cm$^{-3}$ component 
of $\sim$1.8 $\times$ 10$^{-2}$ M$_{\odot}$ was required to explain the data at $\sim$5000 days. Such low density gas is expected in the 
H~II-region interior to the inner ring which likely extends also to larger radii at higher latitudes. Assuming a constant density
between the inner ring radius and out to 1.3 $\times$ this radius, the low density gas would fill a region with an opening angle
of 7.2$^{\circ}$ above and below the ring plane in addition to the gas in the ring. The exact location and extent of this gas is 
not well constrained in our study, but we note that a thick H II-region surrounding the equatorial ring as used in the models of Borkowski
et al. (1997) can at least qualitatively explain the shape of the reverse shock of the ejecta/ring
interaction (Michael et al. 2003). The H II-region used in their model has an extent of $\pm$30$^{\circ}$, 
which is substantially larger than our $\pm$7.2$^{\circ}$. Assuming an opening angle of $\sim$10$^{\circ}$ above and below the
equatorial plane Zhekov et al. (2010) also found a density of $\sim$$10^{2}$ atoms cm$^{-3}$ for the H~II-region based on their X-ray spectra.
There could of course be gas with lower density than
$10^{2}$ atoms cm$^{-3}$ at higher latitudes than in our model. Such gas would probably not contribute much to the
strength of the narrow lines, but may still give rise to a similar structure of the reverse shock as inferred by
Michael et al. (2003, see also Heng et al. 2006). At epochs later than $\sim$5000 days our models underproduce the
emission of most of these lines as expected due to the contribution from the interaction of the SN ejecta with the ring.

\section{Conclusions}
We have presented optical and near-IR line light curves of the inner circumstellar ring of SN 1987A covering the 
entire era from $\sim$300 days when the ring first became visible at these wavelengths to
$\sim$5000 days when the collision between the SN ejecta and the CSM ring started to significantly
affect these fluxes. We modeled the line fluxes with a photoionization code which follows the recombination and 
cooling of the gas in the ring after the initial ionization by the SN flash. From this we draw the following conclusions.

The absolute line fluxes of H$\alpha$ and H$\beta$, and the light curves for most of the lines studied here are modeled well 
with a combination of three density components 1$\times$10$^{3}$, $\sim$3$\times$10$^{3}$ and 3$\times$10$^{4}$ atoms cm$^{-3}$. The 
total mass of the ionized gas found was $\sim$5.8 $\times$ 10$^{-2}$ M$_{\odot}$ with about 20\%, 60\% and 20\% of the mass in the 
three different components, respectively. Complementing our optical/near-IR abundances with estimates of Lundqvist \& Fransson (1996) 
based on UV lines we found an He/H-ratio (by number of atoms) of 0.17 $\pm$ 0.06 and an N/O-ratio of 1.5 $\pm$ 0.7. 
This helium abundance is roughly 30\% lower than previously estimated but still $\sim$2.0 times the
solar and the LMC values. Our estimated
N/O-ratio is higher than typically found for the B-type supergiants in the LMC. However, a very high N/O-ratio has also been found for Sher
25 (and its circumstellar nebula) which has sometimes been suggested to be a twin of the progenitor of SN 1987A. The total
(C+N+O)/(H+He) abundance roughly 1.6 times its LMC value was obtained which is as expected assuming the abundances of metals to be
enhanced by the same factor ($\sim$2.0) as He compared to their LMC values. Our iron abundance based on optical and near-IR
lines is 0.20 $\pm$ 0.11 times solar which is within the range of the estimates for the LMC and therefore does not indicate
a strong depletion onto dust grains in the ring material.

An additional $10^{2}$ atoms cm$^{-3}$ component of $\sim$1.8 $\times$ 10$^{-2}$ M$_{\odot}$ was required to explain the fluxes of a 
few high ionization lines studied at the late times. Such low density gas is expected in the H II-region interior to the inner ring 
which likely extends also to larger radii at higher latitudes. For a few lines we also presented light curves at epochs later than 
$\sim$5000 days and found our models to underproduce the emission as expected due to the contribution from the interaction of the 
SN ejecta with the ring. The re-ionization of this gas by the X-rays from the interaction can therefore be expected to make the ring
glow again especially in many high-ionization narrow UV-lines making it an interesting target for future ground and space based
observations.

\acknowledgments

We thank an anonymous referee for useful comments and Linda Smith, Roberto Terlevich, Robert Cumming, and Stephen Smartt for helpful discussions.
This paper is based on observations made with the AAT and with ESO telescopes at the La Silla Paranal Observatory under programme IDs
66.D-0589, 70.D-0379, 074.D-0761, 078.D-0521, and 080.D-0727. SM acknowledges financial support from the Academy of Finland (project: 8120503),
and PL is grateful for financial support from the Swedish Research Council.

\end{document}